\documentclass[%
 reprint,
superscriptaddress,
 amsmath,amssymb,
 aps,
]{revtex4-2}

\usepackage{graphicx}
\usepackage{dcolumn}
\usepackage{bm}
\usepackage{color}

\begin{document}

\preprint{APS/123-QED}

\title{Electrically Switchable Circular Photogalvanic Effect in Methylammonium Lead Iodide Microcrystals}

\author{Yuqing Zhu}
\author{Ziyi Song}
\author{Rodrigo Becerra Silva}
\author{Bob Minyu Wang}
\author{Henry Clark Travaglini}
\affiliation{Department of Physics and Astronomy, University of California, Davis}
\author{Andrew C Grieder}
\author{Yuan Ping}
\affiliation{Department of Materials Science and Engineering, University of Wisconsin-Madison}
\author{Liang Z. Tan}
\affiliation{Molecular Foundry, Lawrence Berkeley National Lab}
\author{Dong Yu}
 \email{yu@physics.ucdavis.edu}
\affiliation{Department of Physics and Astronomy, University of California, Davis}


\begin{abstract}
We investigate the circular photogalvanic effect (CPGE) in single-crystalline methylammonium lead iodide microcrystals under a static electric field. The external electric field can enhance the magnitude of the helicity dependent photocurrent (HDPC) by two orders of magnitude and flip its sign, which we attribute to magnetic shift currents induced by the Rashba-Edelstein effect. This HDPC induced by the static electric field may be viewed as an unusually strong third-order photoresponse, which produces a current two orders of magnitude larger than second-order injection current. Furthermore, the HDPC is highly nonlocal and can be created by photoexcitation out of the device channel, indicating a spin diffusion length up to 50 $\mu$m at 78 K. 
\end{abstract}

\maketitle



Hybrid lead halide perovskites based solar cells have demonstrated high power conversion efficiency, thanks to their unusually long charge carrier lifetime. The mechanism for the slow recombination is still under intense debate. One possible explanation is that the strong spin-orbit coupling (SOC) in these materials \cite{niesner2016giant} causes indirect spin-polarized subbands via Rashba effects \cite{kepenekian2017rashba, zheng2015rashba}. Materials with strong SOC also have potentials for realizing efficient charge-spin conversion, key to spintronic applications \cite{kepenekian2015rashba}. Despite its importance, the origin of the Rashba effects in halide perovskites remains unclear. In particular, the methylammonium lead iodide (MAPbI\textsubscript{3}) and bromide (MAPbBr\textsubscript{3}) crystals obey inversion symmetry in both their tetragonal and orthorhombic phases, forbidding Rasbha effects in the bulk. However, it has been proposed that a 'dynamic' Rashba splitting in the bulk can be induced by symmetry-breaking thermal fluctuation \cite{etienne2016dynamical, niesner2018structural, schlipf2021dynamic}. On the other hand, substantial reports support that the 'static' surface Rashba splitting can dominate in this system \cite{frohna2018inversion, ryu2020static, huang2021observation}.

\begin{figure*}
\centering
\includegraphics[width=12 cm]{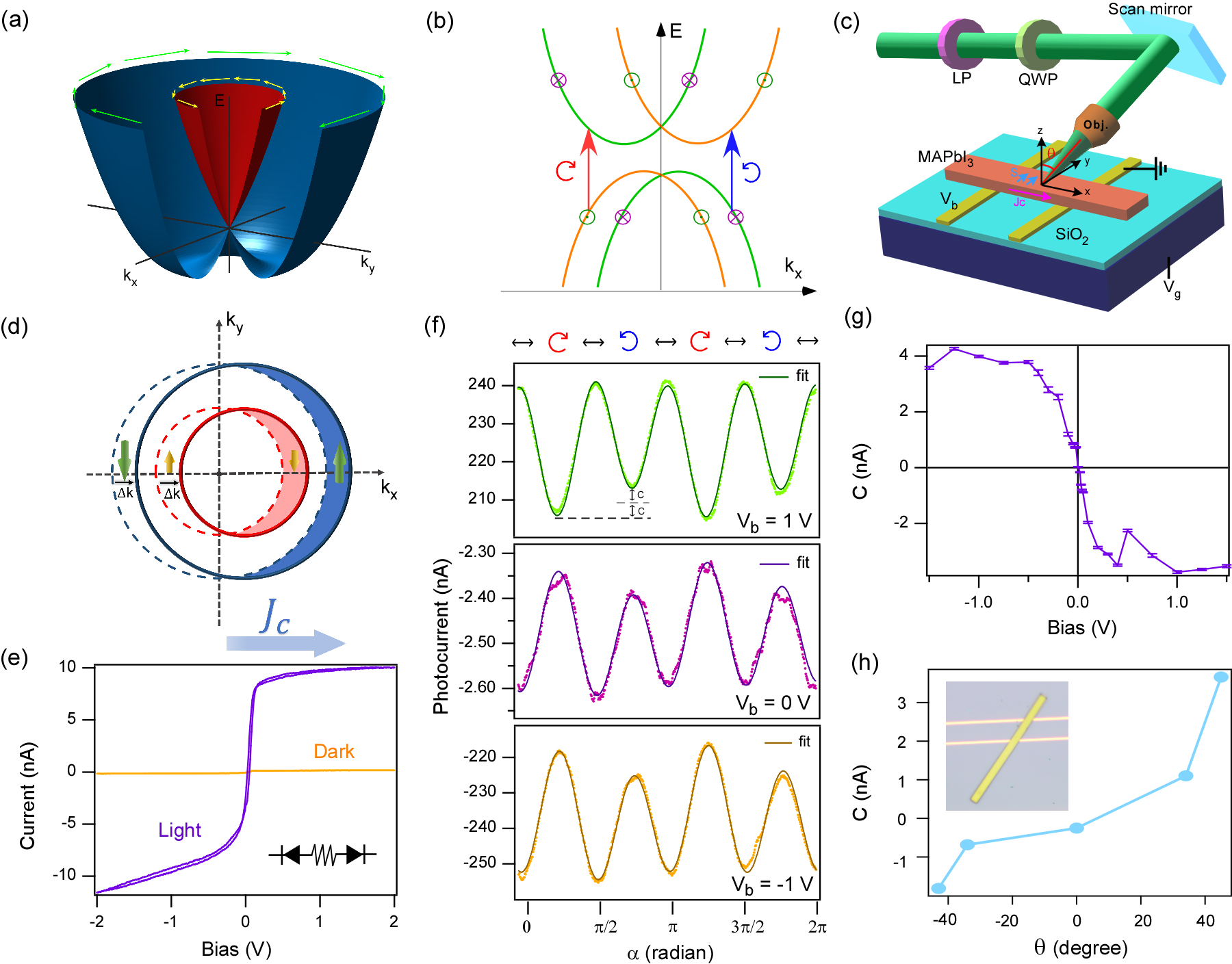}
\caption{\label{fig:bias} Bias switchable CPGE. (a) Illustration of Rashba spin bands. (b) Photon helicity dependent optical transitions in a Rashba system. (c) Experimental setup where a laser beam with its polarization controlled by a rotating QWP is focused on a microbeam FET. A charge current $J_c$ driven by source-drain bias creates surface spin polarization ($S$) by the Edelstein effect. (d) Illustration of charge current-induced spin polarization by the Edelstein effect. (e) Current-voltage curves in dark and under uniform illumination with an intensity about 3.5 mW/cm$^2$. Inset: a back-to-back diode circuit model to understand the current saturation. (f) Photocurrent as a function of the QWP angle $\alpha$ at various biases. (g) $C$ vs $V_b$, with the laser focused at reversely biased contact where photoresponse is strongest. (h) $C$ vs $\theta$ taken at $V_b = -1 V$ in a different device. Inset: an optical image of a typical device. The gap between electrodes is 20 $\mu$m. All data were taken at 78 K. The laser with wavelength of 650 nm and power of 1 $\mu$W was incident at $\theta$ = 45$^{\circ}$ in (f) and (g). The same wavelength and power were used for (h).}
\end{figure*}

The circular photogalvanic effect (CPGE) experiment measures how photocurrent depends on the circular polarization of the incident light [Fig. \ref{fig:bias}(b)]. Circularly polarized light can selectively induce interband optical transition with unbalanced charge current. This technique has been used to probe the spin textures in quantum wells \cite{ganichev2003spin}, topological insulators \cite{mciver2012control}, and 2D electron gas in oxides \cite{wang2022circular}. The CPGE has also been used recently to study the Rashba effects in both 3D and 2D halide perovskites \cite{niesner2018structural,liu2020circular}. However, the previous CPGE studies in these materials were performed at zero electrical bias and with limited spatial resolution. Here we investigate Rashba effects with spatially and energetically resolved CPGE, under an external static electric field and at various temperatures. We show that the electric field has a high impact on the CPGE. 


Devices composed of individual pristine micrometer-sized single crystals of MAPbI\textsubscript{3} were used to mitigate the impacts of grain boundaries as in polycrystalline thin films, while still maintaining a relatively large surface-to-volume ratio. Single crystalline microstructures of MAPbI\textsubscript{3} were synthesized following a dissolution and recrystallization process~\cite{fu2015solution}. Microbeams and microplates with thickness of about 1 $\mu$m and length up to 100 $\mu$m were produced with well defined facets and smooth surfaces (Fig. S1 in Supplemental Materials). The microcrystals were then mechanically transferred to pre-patterned Au electrodes on SiO\textsubscript{2} coated Si substrates without exposure to the detrimental lithographic process. Devices composed of freshly grown samples were transferred into an optical cryostat (Janis ST-500) and pumped to about 10$^{-6}$ Torr. A tunable laser (NKT Photonics)  with wavelength variable in 500-800 nm was focused by a 10x objective lens to a diameter about 3 $\mu$m, incident at $\theta$ = 45$^{\circ}$ and perpendicular to the device channel. The oblique incidence was achieved using a tilted sample stage inside the optical cryostat and the laser spot was elongated into an ellipse. Circular polarization was achieved by passing a linearly polarized beam through a rotating achromatic quarter waveplate (QWP, Thorlabs). The output polarization is a function of the angle between the QWP fast axis and the incident linear polarization ($\alpha$), continuously changing between left circular polarization (LCP), to linear polarization (LP), and to right circular polarization (RCP) as the QWP is rotated. A schematic drawing of the experimental setup is shown in Fig. \ref{fig:bias}(c). 

We first present our CPGE measurement results at 78 K. The device is insulating in the dark but becomes conductive under light. The current saturates at source-drain bias ($V_b$) over 0.2 V [Fig. \ref{fig:bias}(e)], which is caused by the Schottky contacts and can be understood with the back-to-back diode circuit model. The photocurrent created by the laser focused near the contact clearly differs between LCP and RCP, as shown in Fig. \ref{fig:bias}(f). The photocurrent can be well fit by,

\begin{equation}
    \centering
    I=C\sin(2\alpha)+L_1\sin(4\alpha)+L_2\cos(4\alpha)+D
    \label{CPGE Eqn}
\end{equation}

\noindent where $C$ represents the amplitude of CPGE, $L_1$ the linear polarization dependent effects, $L_2$ the reflectance difference at $s$ and $p$ polarizations, and $D$ the polarization independent contributions. All these parameters are linear with the laser power as shown in Fig. S2. $C$ vanishes at normal incidence and flips sign with negative $\theta$ [Fig. \ref{fig:bias}(h)], indicating the injection of in-plane spins is needed to produce CPGE. Interestingly, the magnitude of helicity dependent photocurrent (HDPC) is greatly enhanced under bias and its sign is flipped at the reversed bias [Fig. \ref{fig:bias}(f-g)]. $C$ is non-zero but small about 0.02 nA at zero bias. It then quickly increases to over 4 nA under $V_b$ = 0.5 V and saturates similar to the $I-V_b$ curve in Fig. \ref{fig:bias}(e). The magnitude of $C$ per incident power in our devices is found to be about 0.02 to 4 mA/W, depending on the bias. These values are about one to three orders of magnitude larger than previously reported in mm-sized single crystal MAPbI\textsubscript{3} \cite{niesner2018structural} at zero bias.

The bias enhanced CPGE has been reported in Si nanowires \cite{dhara2015voltage} and MoS\textsubscript{2} monolayers \cite{liu2018electrical}, attributed to electric field induced symmetry breaking and enhanced charge separation at contact, respectively. These mechanisms are not applicable to our halide perovskite system. First, our MAPbI\textsubscript{3} microcrystals do not contain zigzag chains as in Si nanowires grown along the $\langle 110 \rangle$ direction. Second, though the charge separation at the reversely biased contact is also enhanced by bias in our device resulting in larger $C$, $D$, $L_1$, and $L_2$, the sign flipping of $C$ at opposite bias is unexpected, since the spin-momentum chirality at the top surface is determined by the out-of-plane electric field independent of bias. The third-order photocurrents, including jerk, injection, and shift, are produced by the combination of static and optical electric fields and its polarity can be flipped by the static field \cite{fregoso2019bulk}. But detailed analysis shows that all these third-order photocurrents are negligible in the Rashba system in our measurement geometry (see Supplemental Materials).

\begin{figure*}[hbt!]
\centering
\includegraphics[width=17cm]{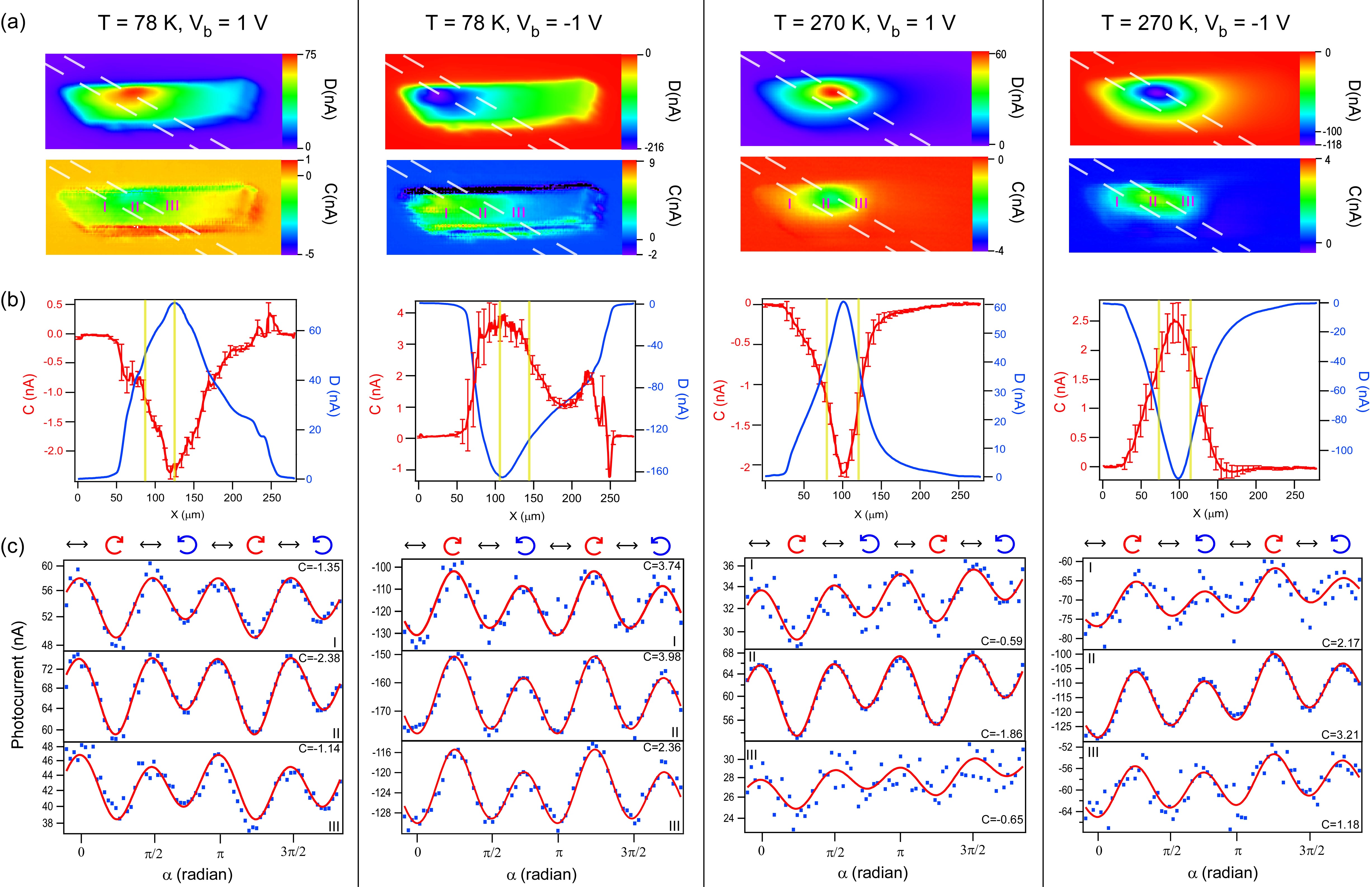}
\caption{\label{fig:map} Spatially resolved CPGE at 78 and 270 K with $V_b$ = $\pm$ 1 V . (a) $D$ and $C$ maps of a microplate device. Dashed lines indicate the contact positions. (b) Cross sections of $C$ and $D$ along the channel. Golden vertical lines denote the contact positions. (c) Photocurrent as a function of $\alpha$ at positions I-III denoted in the maps. The laser with wavelength of 532 nm and power of 1 $\mu$W was incident at $\theta$ = 45$^{\circ}$.}
\end{figure*}

We attribute the bias switchable HDPC to the Rashba-Edelstein effect (REE) \cite{edelstein1990spin}, where the applied electric field shifts the Fermi surface and leads to spin polarization at the surface [Fig. \ref{fig:bias}(d)]. The spin polarization then lifts the Rashba band degeneracies, breaks time-reversal symmetry, and leads to a strong magnetic shift current (MSC). The bias switching of the $C$ sign can then be understood by the spin polarization reversal at the surface. REE induced spin-charge conversion has been demonstrated in topological insulator $\alpha$-Sn \cite{rojas2016spin} and monolayer transition metal dichalcogenides \cite{shao2016strong}. Recently, REE induced spin to charge conversion has also been reported in MAPbBr\textsubscript{3} via spin injection from a ferromagnetic interface \cite{sun2019surface}. We note that spin Hall effect (SHE) can also result in spin polarization through bulk transverse spin current \cite{trier2019electric}. However, SHE has not been demonstrated in MAPbI\textsubscript{3}. MSC is a dc charge current known to be induced by circularly polarized light in magnetic systems~\cite{chen_basic_2022,  wang_electrically_2020,watanabe_photocurrent_2021, pi2023magnetic} as part of broader work on $\mathcal{PT}$-symmetric systems~\cite{zhang_switchable_2019, fei_giant_2020, fei_pt-symmetry-enabled_2021, xue_valley_2023, liu_switchable_2023}, but has not been previously considered in non-magnetic systems with photoinduced spin populations such as halide perovskites. The large $C$ value under bias indicates that REE and MSC can produce stronger CPGE than the Rashba effect alone. 


Spatially resolved CPGE measurements are then performed by scanning photocurrent microscopy (SPCM), a powerful experimental technique that provides spatially resolved photocurrent mapping and insights on carrier transport~\cite{fu2011electrothermal, graham2013scanning, xiao2016photocurrent, wang2017dynamic}. To investigate CPGE, photocurrent images are taken at various $\alpha$'s. $C$ and $D$ are then obtained by fitting the $\alpha$ dependent photocurrent at each pixel of the photocurrent images by Eqn. (\ref{CPGE Eqn}) (more details can be found in ref. \cite{wang2023spatially}). The $C$ and $D$ maps and their distributions along the microplate longitudinal axis at $V_b = \pm 1$ V at 78 K and 270 K are shown in Fig.~\ref{fig:map}. $D$ reaches maximum at the reversely biased contact. At 78 K, the $D$ value decays slowly when the laser spot is scanned away from the contact out of the device channel; a long carrier diffusion length of $100 \pm 20 \ \mu$m is extracted from exponential fitting. In comparison, the photocurrent decay length is much shorter, about $33 \pm 4 \ \mu$m at 270 K. All these are consistent with our previous work \cite{xiao2016photocurrent, mcclintock2020temperature}. Surprisingly, at 78 K, $C$ extends out of the device channel as well with a decay length about  $50 \pm 3 \mu$m. The bias can also flip the sign of $C$ outside the channel. The large and oscillating $C$ values near the edges of the microplate is an artifact due to the rapid change of $D$, evidenced by scattered $\alpha$ dependence at those positions (Fig. S3). The $\alpha$ dependence away from the edges can be fit very well by Eqn. \ref{CPGE Eqn}, as shown in Fig. \ref{fig:map} (c). $L1$ and $L2$ maps also show nonlocal behaviors at 78 K (Fig. S4). The nonlocal CPGE is also observed consistently in various devices (Fig. S5). The decay lengths of $D$ and $C$ drop quickly at 270 K. The uncertainty of $C$ is large outside the channel at 270 K as evidenced by the scattered $\alpha$ dependence plots. It is unclear whether we have any CPGE out of channel at 270 K. 

This bias switchable HDPC from the outside of the device channel can also be understood by the surface spin generation within the channel by charge current via REE, followed by the diffusion of these spins out of the channel. The highly nonlocal $C$ at 78 K indicates a long spin diffusion length. The reported electron spin relaxation times in MAPbI\textsubscript{3} vary in a wide range, from 7 ps to 37 ns, depending on temperatures and the materials \cite{giovanni2015highly, odenthal2017spin, kirstein2022spin, xu2024spin}. A spin diffusion length of 60 nm at room temperature has been reported in MAPbCl\textsubscript{3-x}I\textsubscript{x} thin films \cite{yang2019unexpected}. For a diffusion length of 50 $\mu$m at 78 K, we estimate a carrier mobility of 10$^5$ cm$^2$/Vs, if using spin relaxation time of 37 ns. This mobility value is on the higher end of the expected values~\cite{frost2017calculating} and may indicate the contributions from highly mobile excitons. At low temperatures, excitons are expected to form and their diffusion and spin relaxation are currently not well understood. We have demonstrated experimentally that excitons can have significantly higher mobilities than free carriers, thanks to their dipolar nature, which suppresses phonon scattering ~\cite{mcclintock2020temperature, tang2021transport, mcclintock2022highly}. These highly mobile excitons are spin polarized under circular polarization and can diffuse out of channel, creating the observed non-local CPGE. 

Ferroelectric domains have been reported in MAPbI\textsubscript{3} \cite{frost2014atomistic, garten2019existence}. Such ferroelectric polarization has been theoretically proposed to create and enhance Rashba splitting in the bulk \cite{kim2014switchable, leppert2016electric}. However, electric field induced ferroelectric polarization unlikely accounts for our observed bias switchable CPGE, because our CPGE switching is also observed when the photoexcitation is out of the device channel, where the external electric field vanishes. In addition, CPGE induced by the bulk Rashba effect with the external electric field along the channel is not expected to follow the incident angle dependence as observed. 


\begin{figure}[t]
\centering
\includegraphics[width=8.4cm]{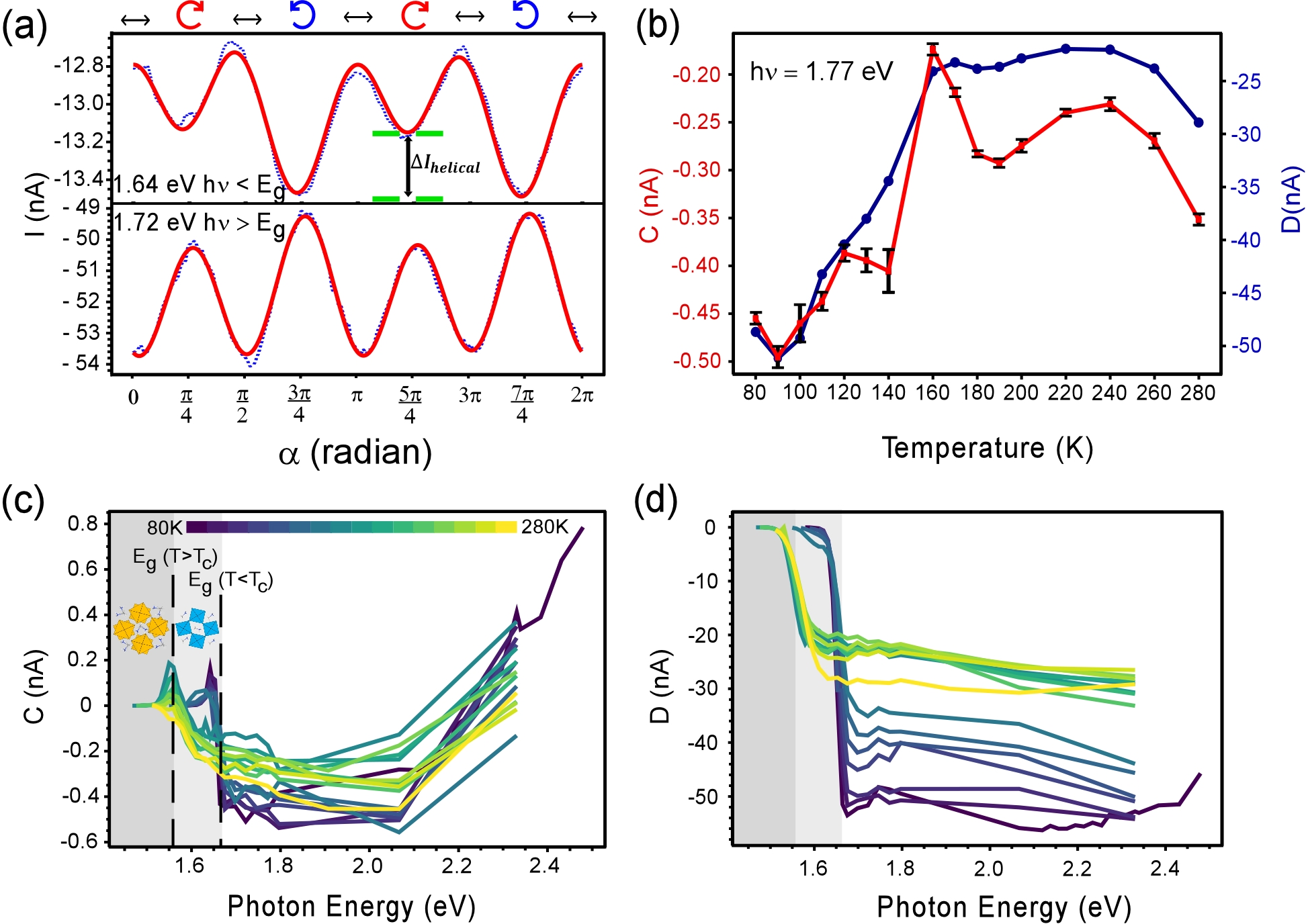}
\caption{\label{fig:T} Photon energy dependent CPGE at various temperatures. (a) Photocurrent as a function of $\alpha$ with photon energy below and above bandgap at 80 K. (b) Temperature dependent $C$ and $D$ at photon energy of 1.77 eV. (c, d) Photon energy dependent $C$ and $D$ at various temperatures. The crystal undergoes an orthorhombic to tetragonal phase transition above $T_c \approx$ 150 K associated with a abrupt bandgap decrease as indicated. The laser of power of 1 $\mu$W is incident at $\theta$ = 45$^{\circ}$. The bias is fixed at -1 V.}
\end{figure}

The photon energy effects on CPGE at various temperatures are shown in Fig. \ref{fig:T}. $D$ sharply increases above the bandgap [Fig. \ref{fig:T}(d)]. The onset photon energy changes abruptly when the crystal undergoes transition from the orthorhombic to the tetragonal phase at $T_c \approx$ 150 K. A small bump at 1.67 eV can be observed at low temperatures, most obvious at 80 K, hinting an excitonic peak \cite{phuong2016free, mcclintock2022highly}. The magnitude of $D$ stays about the same above $T_c$ and increases with decreasing temperature below $T_c$ [Fig. \ref{fig:T}(b)]. As shown in Fig. \ref{fig:T}(c), the wavelength dependence of $C$ at different temperatures largely follows a similar trend. Above the bandgap, its magnitude first increases and then decreases as the photon energy increases. Its sign flips when the photon energy is over 2.2-2.3 eV at low temperatures. Most strikingly, $C$ flips sign when the photon energy is below the bandgap [Fig. \ref{fig:T}(a) and Fig. S6]. A narrow peak with a width about thermal energy $K_BT$ is clearly seen at T $<$ 200 K. After the phase transition, the $C$ peak position shifts accordingly along with the bandgap. At T $>$ 200 K, the $C$ peak significantly drops in magnitude and eventually vanishes. $C$ also flips sign below bandgap when the bias is reversed. The 78 K $C$ map obtained at photon energy below bandgap also shows nonlocal behavior but with a shorter decay length as shown in Fig. S7.  

The magnitude of $C$ increases at lower temperatures, different from the previous results measured in millimeter-sized single crystals \cite{niesner2018structural}, where the CPGE increases with temperature. Therefore, the observed CPGE in our microcrystals is unlikely caused by the thermal fluctuation induced dynamic Rashba effect. The interband CPGE spectrum is broad and largely follows the theoretical calculations \cite{li2016circular, liu2020circular}. We also carried out first-principles calculations of the CPGE tensor and the calculation results are qualitatively consistent with the experimental observation (see more detail later). The exact mechanism for the observed below-gap $C$ sign flipping is unclear and we propose two possible interpretations below. First, the flipping may be caused by the absorption depth change. Below bandgap, the light can penetrate through the microplate and be directly absorbed near the bottom surface with spin polarization opposite to the top surface. The overall HDPC is the sum of the top and bottom surface contributions and may flip sign, if the bottom surface dominates because of its vicinity to the bottom contacts. However, this does not explain the disappearance of $C$ peaks at higher temperatures. Second, a similar $C$ sign flipping below bandgap has been observed in 2D halide perovskites and attributed to spin-galvanic effect (SGE) of the ionized spin-polarized excitons \cite{liu2020circular}. MAPbI\textsubscript{3} has an exciton binding energy of 16 meV determined by magnetic field dependent optical absorption~\cite{miyata2015direct}. Free carriers and excitons likely coexist at low temperatures in this material \cite{tang2021transport}. Excitons and free carriers may lead to opposite HDPC because of their distinct CPGE mechanisms \cite{liu2020circular}.  At higher temperatures the below-gap $C$ peak vanishes, when the thermal energy forbids exciton formation. 
 
To better understand the origin of CPGE and its wavelength dependence, we have performed first-principles calculations of band structures and CPGE tensor ($\beta_{ij}$) for a MAPbI\textsubscript{3} slab. An asymmetric slab was constructed to ensure inversion symmetry breaking. This results in Rashba band splitting in the conduction and valence band edges, as shown in Supplemental Materials Fig. S8. The slab is terminated along the (001) surface as previously shown to be one of the most stable surface orientations~\cite{wang2015}. The ionic positions in the slab are fully relaxed before calculating $\beta_{ij}$ and in turn the second order injection current, see Supplemental Materials for more details.  We also computed the current dependence on incident light propagation angle and photon frequency, which qualitatively agree with experiments, as shown in Fig. S9. The nature of the angle dependence is a consequence of the symmetry of the CPGE tensor. We find $\beta_{xy}=-\beta_{yx}$ and almost zero for all other components, consistent with a crystal having $C_{4v}$ point group symmetry~\cite{Cle2021}. While the relaxed slab does not strictly have this symmetry, we find the inorganic sublattice is close to $C_{4v}$ point group. This also explains the zero crossing in Fig.~\ref{fig:bias}(h) when the light is normal to the surface (0$^\circ$). At 45$^\circ$, we calculate the second-order injection current of 0.028 nA, which agrees well with the zero-bias value measured in experiments of 0.02 nA. The width of the calculated CPGE peak is about 1.5 eV, broader than that measured experimentally (0.6 eV). The deviation can be caused by the surface chemical termination difference, the omission of scattering-induced ballistic currents in the DFT calculation, and the crossover between the magnetic shift current and the injection current.

On the application of bias, the REE introduces time-reversal symmetry breaking terms into the Rashba Hamiltonian which drastically modifies the photoresponse. Our DFT calculations reveal that the out-of-plane transition dipoles of a MAPbI\textsubscript{3} slab are especially strong near the Rashba band degeneracies, due to the surface potential. Ordinarily, these transitions do not affect the total absorption or the injection current by much because of the small density of states near the Rashba band degeneracies. However, under time-reversal symmetry breaking of the REE, these band degeneracies are gapped and replaced by highly-divergent van Hove singularities, leading to large MSC at these parts of the Brillouin zone. Our Rashba model calculations show that MSC can typically be two orders of magnitude stronger than injection currents for the parameters of our system (see Supplemental Materials). The calculated MSC also vanishes at normal incidence in agreement with the experimental result.   

These results suggest that transient spin populations may provide a novel route to enhancing photocurrents in other Rashba-type materials. As we have shown here, this mechanism does not require any ground state magnetism, and is capable of producing photocurrents two orders of magnitude larger than other bulk photovoltaic effect mechanisms. While we have demonstrated this for a system where REE is confined to the surface, the same may be possible for bulk Rashba-type materials, where we expect even larger total currents collected over the bulk of the sample. 

In summary, we showed that an external static electric field can enhance the CPGE photocurrent by two orders of magnitude and flip its sign. We proposed a novel mechanism involving the third order MSC, as a result of the REE induced spin polarization at the surface. We also observed that the HDPC magnitude increases at lower temperatures, implying the CPGE in our microcrystals is unlikely caused by the thermally induced dynamic Rashba effect. Furthermore, the CPGE photocurrent is highly nonlocal, indicating a spin diffusion length of 50 $\mu$m at 78 K. While the interband CPGE photocurrent shows a broad spectral distribution above the bandgap largely consistent with the DFT calculations, the CPGE photocurrent below the bandgap flips sign and shows a peak with width resembling the thermal energy at low temperatures. These spatially and spectrally resolved photocurrent results can be related to the formation of highly mobile and spin-polarized excitons.  

\begin{acknowledgments}
We thank Andrew Parsons, Qiaochu Wang, Eric Beery, Pedro De Oliveira, and Mykyta Dementyev for their assistance in the experiments. Z. Song and R.B. Silva made equal contributions to this work. This work was supported by the U.S. National Science Foundation Grants DMR-2209884. Part of this study was performed at the UC Davis Center for Nano and Micro Manufacturing (CNM2). Theory and simulation were supported by the Computational Materials Sciences Program funded by the US Department of Energy, Office of Science, Basic Energy Sciences, Materials Sciences and Engineering Division.  Additional effective Hamiltonian modelling was supported by a user project at the Molecular Foundry was supported by the Office of Science, Office of Basic Energy Sciences, of the U.S. Department of Energy under Contract No. DE-AC02-05CH11231. This research used resources of the National Energy Research Scientific Computing Center, a DOE Office of Science User Facility supported by the Office of Science of the U.S. Department of Energy under Contract No. DE-AC02-05CH11231.
\end{acknowledgments}

\nocite{*}


\begin{thebibliography}{56}%
\makeatletter
\providecommand \@ifxundefined [1]{%
 \@ifx{#1\undefined}
}%
\providecommand \@ifnum [1]{%
 \ifnum #1\expandafter \@firstoftwo
 \else \expandafter \@secondoftwo
 \fi
}%
\providecommand \@ifx [1]{%
 \ifx #1\expandafter \@firstoftwo
 \else \expandafter \@secondoftwo
 \fi
}%
\providecommand \natexlab [1]{#1}%
\providecommand \enquote  [1]{``#1''}%
\providecommand \bibnamefont  [1]{#1}%
\providecommand \bibfnamefont [1]{#1}%
\providecommand \citenamefont [1]{#1}%
\providecommand \href@noop [0]{\@secondoftwo}%
\providecommand \href [0]{\begingroup \@sanitize@url \@href}%
\providecommand \@href[1]{\@@startlink{#1}\@@href}%
\providecommand \@@href[1]{\endgroup#1\@@endlink}%
\providecommand \@sanitize@url [0]{\catcode `\\12\catcode `\$12\catcode `\&12\catcode `\#12\catcode `\^12\catcode `\_12\catcode `\%12\relax}%
\providecommand \@@startlink[1]{}%
\providecommand \@@endlink[0]{}%
\providecommand \url  [0]{\begingroup\@sanitize@url \@url }%
\providecommand \@url [1]{\endgroup\@href {#1}{\urlprefix }}%
\providecommand \urlprefix  [0]{URL }%
\providecommand \Eprint [0]{\href }%
\providecommand \doibase [0]{https://doi.org/}%
\providecommand \selectlanguage [0]{\@gobble}%
\providecommand \bibinfo  [0]{\@secondoftwo}%
\providecommand \bibfield  [0]{\@secondoftwo}%
\providecommand \translation [1]{[#1]}%
\providecommand \BibitemOpen [0]{}%
\providecommand \bibitemStop [0]{}%
\providecommand \bibitemNoStop [0]{.\EOS\space}%
\providecommand \EOS [0]{\spacefactor3000\relax}%
\providecommand \BibitemShut  [1]{\csname bibitem#1\endcsname}%
\let\auto@bib@innerbib\@empty
\bibitem [{\citenamefont {Niesner}\ \emph {et~al.}(2016)\citenamefont {Niesner}, \citenamefont {Wilhelm}, \citenamefont {Levchuk}, \citenamefont {Osvet}, \citenamefont {Shrestha}, \citenamefont {Batentschuk}, \citenamefont {Brabec},\ and\ \citenamefont {Fauster}}]{niesner2016giant}%
  \BibitemOpen
  \bibfield  {author} {\bibinfo {author} {\bibfnamefont {D.}~\bibnamefont {Niesner}}, \bibinfo {author} {\bibfnamefont {M.}~\bibnamefont {Wilhelm}}, \bibinfo {author} {\bibfnamefont {I.}~\bibnamefont {Levchuk}}, \bibinfo {author} {\bibfnamefont {A.}~\bibnamefont {Osvet}}, \bibinfo {author} {\bibfnamefont {S.}~\bibnamefont {Shrestha}}, \bibinfo {author} {\bibfnamefont {M.}~\bibnamefont {Batentschuk}}, \bibinfo {author} {\bibfnamefont {C.}~\bibnamefont {Brabec}},\ and\ \bibinfo {author} {\bibfnamefont {T.}~\bibnamefont {Fauster}},\ }\bibfield  {title} {\bibinfo {title} {{Giant rashba splitting in CH3NH3PbBr3 organic-inorganic perovskite}},\ }\href@noop {} {\bibfield  {journal} {\bibinfo  {journal} {Phys. Rev. Lett.}\ }\textbf {\bibinfo {volume} {117}},\ \bibinfo {pages} {126401} (\bibinfo {year} {2016})}\BibitemShut {NoStop}%
\bibitem [{\citenamefont {Kepenekian}\ and\ \citenamefont {Even}(2017)}]{kepenekian2017rashba}%
  \BibitemOpen
  \bibfield  {author} {\bibinfo {author} {\bibfnamefont {M.}~\bibnamefont {Kepenekian}}\ and\ \bibinfo {author} {\bibfnamefont {J.}~\bibnamefont {Even}},\ }\bibfield  {title} {\bibinfo {title} {Rashba and dresselhaus couplings in halide perovskites: Accomplishments and opportunities for spintronics and spin--orbitronics},\ }\href@noop {} {\bibfield  {journal} {\bibinfo  {journal} {J. Phys. Chem. Lett.}\ }\textbf {\bibinfo {volume} {8}},\ \bibinfo {pages} {3362} (\bibinfo {year} {2017})}\BibitemShut {NoStop}%
\bibitem [{\citenamefont {Zheng}\ \emph {et~al.}(2015)\citenamefont {Zheng}, \citenamefont {Tan}, \citenamefont {Liu},\ and\ \citenamefont {Rappe}}]{zheng2015rashba}%
  \BibitemOpen
  \bibfield  {author} {\bibinfo {author} {\bibfnamefont {F.}~\bibnamefont {Zheng}}, \bibinfo {author} {\bibfnamefont {L.~Z.}\ \bibnamefont {Tan}}, \bibinfo {author} {\bibfnamefont {S.}~\bibnamefont {Liu}},\ and\ \bibinfo {author} {\bibfnamefont {A.~M.}\ \bibnamefont {Rappe}},\ }\bibfield  {title} {\bibinfo {title} {{Rashba spin--orbit coupling enhanced carrier lifetime in CH3NH3PbI3}},\ }\href@noop {} {\bibfield  {journal} {\bibinfo  {journal} {Nano Lett.}\ }\textbf {\bibinfo {volume} {15}},\ \bibinfo {pages} {7794} (\bibinfo {year} {2015})}\BibitemShut {NoStop}%
\bibitem [{\citenamefont {Kepenekian}\ \emph {et~al.}(2015)\citenamefont {Kepenekian}, \citenamefont {Robles}, \citenamefont {Katan}, \citenamefont {Sapori}, \citenamefont {Pedesseau},\ and\ \citenamefont {Even}}]{kepenekian2015rashba}%
  \BibitemOpen
  \bibfield  {author} {\bibinfo {author} {\bibfnamefont {M.}~\bibnamefont {Kepenekian}}, \bibinfo {author} {\bibfnamefont {R.}~\bibnamefont {Robles}}, \bibinfo {author} {\bibfnamefont {C.}~\bibnamefont {Katan}}, \bibinfo {author} {\bibfnamefont {D.}~\bibnamefont {Sapori}}, \bibinfo {author} {\bibfnamefont {L.}~\bibnamefont {Pedesseau}},\ and\ \bibinfo {author} {\bibfnamefont {J.}~\bibnamefont {Even}},\ }\bibfield  {title} {\bibinfo {title} {Rashba and dresselhaus effects in hybrid organic--inorganic perovskites: from basics to devices},\ }\href@noop {} {\bibfield  {journal} {\bibinfo  {journal} {ACS Nano}\ }\textbf {\bibinfo {volume} {9}},\ \bibinfo {pages} {11557} (\bibinfo {year} {2015})}\BibitemShut {NoStop}%
\bibitem [{\citenamefont {Etienne}\ \emph {et~al.}(2016)\citenamefont {Etienne}, \citenamefont {Mosconi},\ and\ \citenamefont {De~Angelis}}]{etienne2016dynamical}%
  \BibitemOpen
  \bibfield  {author} {\bibinfo {author} {\bibfnamefont {T.}~\bibnamefont {Etienne}}, \bibinfo {author} {\bibfnamefont {E.}~\bibnamefont {Mosconi}},\ and\ \bibinfo {author} {\bibfnamefont {F.}~\bibnamefont {De~Angelis}},\ }\bibfield  {title} {\bibinfo {title} {Dynamical origin of the rashba effect in organohalide lead perovskites: A key to suppressed carrier recombination in perovskite solar cells?},\ }\href@noop {} {\bibfield  {journal} {\bibinfo  {journal} {J. Phys. Chem. Lett.}\ }\textbf {\bibinfo {volume} {7}},\ \bibinfo {pages} {1638} (\bibinfo {year} {2016})}\BibitemShut {NoStop}%
\bibitem [{\citenamefont {Niesner}\ \emph {et~al.}(2018)\citenamefont {Niesner}, \citenamefont {Hauck}, \citenamefont {Shrestha}, \citenamefont {Levchuk}, \citenamefont {Matt}, \citenamefont {Osvet}, \citenamefont {Batentschuk}, \citenamefont {Brabec}, \citenamefont {Weber},\ and\ \citenamefont {Fauster}}]{niesner2018structural}%
  \BibitemOpen
  \bibfield  {author} {\bibinfo {author} {\bibfnamefont {D.}~\bibnamefont {Niesner}}, \bibinfo {author} {\bibfnamefont {M.}~\bibnamefont {Hauck}}, \bibinfo {author} {\bibfnamefont {S.}~\bibnamefont {Shrestha}}, \bibinfo {author} {\bibfnamefont {I.}~\bibnamefont {Levchuk}}, \bibinfo {author} {\bibfnamefont {G.~J.}\ \bibnamefont {Matt}}, \bibinfo {author} {\bibfnamefont {A.}~\bibnamefont {Osvet}}, \bibinfo {author} {\bibfnamefont {M.}~\bibnamefont {Batentschuk}}, \bibinfo {author} {\bibfnamefont {C.}~\bibnamefont {Brabec}}, \bibinfo {author} {\bibfnamefont {H.~B.}\ \bibnamefont {Weber}},\ and\ \bibinfo {author} {\bibfnamefont {T.}~\bibnamefont {Fauster}},\ }\bibfield  {title} {\bibinfo {title} {{Structural fluctuations cause spin-split states in tetragonal (CH3NH3) PbI3 as evidenced by the circular photogalvanic effect}},\ }\href@noop {} {\bibfield  {journal} {\bibinfo  {journal} {Proc. Natl. Acad. Sci.}\ }\textbf {\bibinfo {volume} {115}},\ \bibinfo {pages} {9509} (\bibinfo {year} {2018})}\BibitemShut
  {NoStop}%
\bibitem [{\citenamefont {Schlipf}\ and\ \citenamefont {Giustino}(2021)}]{schlipf2021dynamic}%
  \BibitemOpen
  \bibfield  {author} {\bibinfo {author} {\bibfnamefont {M.}~\bibnamefont {Schlipf}}\ and\ \bibinfo {author} {\bibfnamefont {F.}~\bibnamefont {Giustino}},\ }\bibfield  {title} {\bibinfo {title} {Dynamic rashba-dresselhaus effect},\ }\href@noop {} {\bibfield  {journal} {\bibinfo  {journal} {Phys. Rev. Lett.}\ }\textbf {\bibinfo {volume} {127}},\ \bibinfo {pages} {237601} (\bibinfo {year} {2021})}\BibitemShut {NoStop}%
\bibitem [{\citenamefont {Frohna}\ \emph {et~al.}(2018)\citenamefont {Frohna}, \citenamefont {Deshpande}, \citenamefont {Harter}, \citenamefont {Peng}, \citenamefont {Barker}, \citenamefont {Neaton}, \citenamefont {Louie}, \citenamefont {Bakr}, \citenamefont {Hsieh},\ and\ \citenamefont {Bernardi}}]{frohna2018inversion}%
  \BibitemOpen
  \bibfield  {author} {\bibinfo {author} {\bibfnamefont {K.}~\bibnamefont {Frohna}}, \bibinfo {author} {\bibfnamefont {T.}~\bibnamefont {Deshpande}}, \bibinfo {author} {\bibfnamefont {J.}~\bibnamefont {Harter}}, \bibinfo {author} {\bibfnamefont {W.}~\bibnamefont {Peng}}, \bibinfo {author} {\bibfnamefont {B.~A.}\ \bibnamefont {Barker}}, \bibinfo {author} {\bibfnamefont {J.~B.}\ \bibnamefont {Neaton}}, \bibinfo {author} {\bibfnamefont {S.~G.}\ \bibnamefont {Louie}}, \bibinfo {author} {\bibfnamefont {O.~M.}\ \bibnamefont {Bakr}}, \bibinfo {author} {\bibfnamefont {D.}~\bibnamefont {Hsieh}},\ and\ \bibinfo {author} {\bibfnamefont {M.}~\bibnamefont {Bernardi}},\ }\bibfield  {title} {\bibinfo {title} {Inversion symmetry and bulk rashba effect in methylammonium lead iodide perovskite single crystals},\ }\href@noop {} {\bibfield  {journal} {\bibinfo  {journal} {Nat. Commun.}\ }\textbf {\bibinfo {volume} {9}},\ \bibinfo {pages} {1829} (\bibinfo {year} {2018})}\BibitemShut {NoStop}%
\bibitem [{\citenamefont {Ryu}\ \emph {et~al.}(2020)\citenamefont {Ryu}, \citenamefont {Park}, \citenamefont {McCall}, \citenamefont {Byun}, \citenamefont {Lee}, \citenamefont {Kim}, \citenamefont {Jeong}, \citenamefont {Kim}, \citenamefont {Kanatzidis},\ and\ \citenamefont {Jang}}]{ryu2020static}%
  \BibitemOpen
  \bibfield  {author} {\bibinfo {author} {\bibfnamefont {H.}~\bibnamefont {Ryu}}, \bibinfo {author} {\bibfnamefont {D.~Y.}\ \bibnamefont {Park}}, \bibinfo {author} {\bibfnamefont {K.~M.}\ \bibnamefont {McCall}}, \bibinfo {author} {\bibfnamefont {H.~R.}\ \bibnamefont {Byun}}, \bibinfo {author} {\bibfnamefont {Y.}~\bibnamefont {Lee}}, \bibinfo {author} {\bibfnamefont {T.~J.}\ \bibnamefont {Kim}}, \bibinfo {author} {\bibfnamefont {M.~S.}\ \bibnamefont {Jeong}}, \bibinfo {author} {\bibfnamefont {J.}~\bibnamefont {Kim}}, \bibinfo {author} {\bibfnamefont {M.~G.}\ \bibnamefont {Kanatzidis}},\ and\ \bibinfo {author} {\bibfnamefont {J.~I.}\ \bibnamefont {Jang}},\ }\bibfield  {title} {\bibinfo {title} {{Static Rashba effect by surface reconstruction and photon recycling in the dynamic indirect gap of APbBr3 (A= Cs, CH3NH3) single crystals}},\ }\href@noop {} {\bibfield  {journal} {\bibinfo  {journal} {J. Am. Chem. Soc.}\ }\textbf {\bibinfo {volume} {142}},\ \bibinfo {pages} {21059} (\bibinfo {year} {2020})}\BibitemShut
  {NoStop}%
\bibitem [{\citenamefont {Huang}\ \emph {et~al.}(2021)\citenamefont {Huang}, \citenamefont {Vardeny}, \citenamefont {Wang}, \citenamefont {Ahmad}, \citenamefont {Chanana}, \citenamefont {Vetter}, \citenamefont {Yang}, \citenamefont {Liu}, \citenamefont {Galli}, \citenamefont {Amassian} \emph {et~al.}}]{huang2021observation}%
  \BibitemOpen
  \bibfield  {author} {\bibinfo {author} {\bibfnamefont {Z.}~\bibnamefont {Huang}}, \bibinfo {author} {\bibfnamefont {S.~R.}\ \bibnamefont {Vardeny}}, \bibinfo {author} {\bibfnamefont {T.}~\bibnamefont {Wang}}, \bibinfo {author} {\bibfnamefont {Z.}~\bibnamefont {Ahmad}}, \bibinfo {author} {\bibfnamefont {A.}~\bibnamefont {Chanana}}, \bibinfo {author} {\bibfnamefont {E.}~\bibnamefont {Vetter}}, \bibinfo {author} {\bibfnamefont {S.}~\bibnamefont {Yang}}, \bibinfo {author} {\bibfnamefont {X.}~\bibnamefont {Liu}}, \bibinfo {author} {\bibfnamefont {G.}~\bibnamefont {Galli}}, \bibinfo {author} {\bibfnamefont {A.}~\bibnamefont {Amassian}}, \emph {et~al.},\ }\bibfield  {title} {\bibinfo {title} {{Observation of spatially resolved Rashba states on the surface of CH3NH3PbBr3 single crystals}},\ }\href@noop {} {\bibfield  {journal} {\bibinfo  {journal} {Appl. Phys. Rev.}\ }\textbf {\bibinfo {volume} {8}} (\bibinfo {year} {2021})}\BibitemShut {NoStop}%
\bibitem [{\citenamefont {Ganichev}\ and\ \citenamefont {Prettl}(2003)}]{ganichev2003spin}%
  \BibitemOpen
  \bibfield  {author} {\bibinfo {author} {\bibfnamefont {S.~D.}\ \bibnamefont {Ganichev}}\ and\ \bibinfo {author} {\bibfnamefont {W.}~\bibnamefont {Prettl}},\ }\bibfield  {title} {\bibinfo {title} {Spin photocurrents in quantum wells},\ }\href@noop {} {\bibfield  {journal} {\bibinfo  {journal} {J. Phys.-Condens. Mat.}\ }\textbf {\bibinfo {volume} {15}},\ \bibinfo {pages} {R935} (\bibinfo {year} {2003})}\BibitemShut {NoStop}%
\bibitem [{\citenamefont {McIver}\ \emph {et~al.}(2012)\citenamefont {McIver}, \citenamefont {Hsieh}, \citenamefont {Steinberg}, \citenamefont {Jarillo-Herrero},\ and\ \citenamefont {Gedik}}]{mciver2012control}%
  \BibitemOpen
  \bibfield  {author} {\bibinfo {author} {\bibfnamefont {J.}~\bibnamefont {McIver}}, \bibinfo {author} {\bibfnamefont {D.}~\bibnamefont {Hsieh}}, \bibinfo {author} {\bibfnamefont {H.}~\bibnamefont {Steinberg}}, \bibinfo {author} {\bibfnamefont {P.}~\bibnamefont {Jarillo-Herrero}},\ and\ \bibinfo {author} {\bibfnamefont {N.}~\bibnamefont {Gedik}},\ }\bibfield  {title} {\bibinfo {title} {Control over topological insulator photocurrents with light polarization},\ }\href@noop {} {\bibfield  {journal} {\bibinfo  {journal} {Nat. Nanotechnol.}\ }\textbf {\bibinfo {volume} {7}},\ \bibinfo {pages} {96} (\bibinfo {year} {2012})}\BibitemShut {NoStop}%
\bibitem [{\citenamefont {Wang}\ \emph {et~al.}(2022)\citenamefont {Wang}, \citenamefont {Zhang}, \citenamefont {Zhang}, \citenamefont {Li}, \citenamefont {Luo}, \citenamefont {Wang}, \citenamefont {Jin},\ and\ \citenamefont {Sun}}]{wang2022circular}%
  \BibitemOpen
  \bibfield  {author} {\bibinfo {author} {\bibfnamefont {S.}~\bibnamefont {Wang}}, \bibinfo {author} {\bibfnamefont {H.}~\bibnamefont {Zhang}}, \bibinfo {author} {\bibfnamefont {J.}~\bibnamefont {Zhang}}, \bibinfo {author} {\bibfnamefont {S.}~\bibnamefont {Li}}, \bibinfo {author} {\bibfnamefont {D.}~\bibnamefont {Luo}}, \bibinfo {author} {\bibfnamefont {J.}~\bibnamefont {Wang}}, \bibinfo {author} {\bibfnamefont {K.}~\bibnamefont {Jin}},\ and\ \bibinfo {author} {\bibfnamefont {J.}~\bibnamefont {Sun}},\ }\bibfield  {title} {\bibinfo {title} {Circular photogalvanic effect in oxide two-dimensional electron gases},\ }\href@noop {} {\bibfield  {journal} {\bibinfo  {journal} {Phys. Rev. Lett.}\ }\textbf {\bibinfo {volume} {128}},\ \bibinfo {pages} {187401} (\bibinfo {year} {2022})}\BibitemShut {NoStop}%
\bibitem [{\citenamefont {Liu}\ \emph {et~al.}(2020)\citenamefont {Liu}, \citenamefont {Chanana}, \citenamefont {Huynh}, \citenamefont {Xue}, \citenamefont {Haney}, \citenamefont {Blair}, \citenamefont {Jiang},\ and\ \citenamefont {Vardeny}}]{liu2020circular}%
  \BibitemOpen
  \bibfield  {author} {\bibinfo {author} {\bibfnamefont {X.}~\bibnamefont {Liu}}, \bibinfo {author} {\bibfnamefont {A.}~\bibnamefont {Chanana}}, \bibinfo {author} {\bibfnamefont {U.}~\bibnamefont {Huynh}}, \bibinfo {author} {\bibfnamefont {F.}~\bibnamefont {Xue}}, \bibinfo {author} {\bibfnamefont {P.}~\bibnamefont {Haney}}, \bibinfo {author} {\bibfnamefont {S.}~\bibnamefont {Blair}}, \bibinfo {author} {\bibfnamefont {X.}~\bibnamefont {Jiang}},\ and\ \bibinfo {author} {\bibfnamefont {Z.}~\bibnamefont {Vardeny}},\ }\bibfield  {title} {\bibinfo {title} {{Circular photogalvanic spectroscopy of Rashba splitting in 2D hybrid organic--inorganic perovskite multiple quantum wells}},\ }\href@noop {} {\bibfield  {journal} {\bibinfo  {journal} {Nat. Commun.}\ }\textbf {\bibinfo {volume} {11}},\ \bibinfo {pages} {323} (\bibinfo {year} {2020})}\BibitemShut {NoStop}%
\bibitem [{\citenamefont {Fu}\ \emph {et~al.}(2015)\citenamefont {Fu}, \citenamefont {Meng}, \citenamefont {Rowley}, \citenamefont {Thompson}, \citenamefont {Shearer}, \citenamefont {Ma}, \citenamefont {Hamers}, \citenamefont {Wright},\ and\ \citenamefont {Jin}}]{fu2015solution}%
  \BibitemOpen
  \bibfield  {author} {\bibinfo {author} {\bibfnamefont {Y.}~\bibnamefont {Fu}}, \bibinfo {author} {\bibfnamefont {F.}~\bibnamefont {Meng}}, \bibinfo {author} {\bibfnamefont {M.~B.}\ \bibnamefont {Rowley}}, \bibinfo {author} {\bibfnamefont {B.~J.}\ \bibnamefont {Thompson}}, \bibinfo {author} {\bibfnamefont {M.~J.}\ \bibnamefont {Shearer}}, \bibinfo {author} {\bibfnamefont {D.}~\bibnamefont {Ma}}, \bibinfo {author} {\bibfnamefont {R.~J.}\ \bibnamefont {Hamers}}, \bibinfo {author} {\bibfnamefont {J.~C.}\ \bibnamefont {Wright}},\ and\ \bibinfo {author} {\bibfnamefont {S.}~\bibnamefont {Jin}},\ }\bibfield  {title} {\bibinfo {title} {Solution growth of single crystal methylammonium lead halide perovskite nanostructures for optoelectronic and photovoltaic applications},\ }\href@noop {} {\bibfield  {journal} {\bibinfo  {journal} {J. Am. Chem. Soc.}\ }\textbf {\bibinfo {volume} {137}},\ \bibinfo {pages} {5810} (\bibinfo {year} {2015})}\BibitemShut {NoStop}%
\bibitem [{\citenamefont {Dhara}\ \emph {et~al.}(2015)\citenamefont {Dhara}, \citenamefont {Mele},\ and\ \citenamefont {Agarwal}}]{dhara2015voltage}%
  \BibitemOpen
  \bibfield  {author} {\bibinfo {author} {\bibfnamefont {S.}~\bibnamefont {Dhara}}, \bibinfo {author} {\bibfnamefont {E.~J.}\ \bibnamefont {Mele}},\ and\ \bibinfo {author} {\bibfnamefont {R.}~\bibnamefont {Agarwal}},\ }\bibfield  {title} {\bibinfo {title} {Voltage-tunable circular photogalvanic effect in silicon nanowires},\ }\href@noop {} {\bibfield  {journal} {\bibinfo  {journal} {Science}\ }\textbf {\bibinfo {volume} {349}},\ \bibinfo {pages} {726} (\bibinfo {year} {2015})}\BibitemShut {NoStop}%
\bibitem [{\citenamefont {Liu}\ \emph {et~al.}(2018)\citenamefont {Liu}, \citenamefont {Lenferink}, \citenamefont {Wei}, \citenamefont {Stanev}, \citenamefont {Speiser},\ and\ \citenamefont {Stern}}]{liu2018electrical}%
  \BibitemOpen
  \bibfield  {author} {\bibinfo {author} {\bibfnamefont {L.}~\bibnamefont {Liu}}, \bibinfo {author} {\bibfnamefont {E.~J.}\ \bibnamefont {Lenferink}}, \bibinfo {author} {\bibfnamefont {G.}~\bibnamefont {Wei}}, \bibinfo {author} {\bibfnamefont {T.~K.}\ \bibnamefont {Stanev}}, \bibinfo {author} {\bibfnamefont {N.}~\bibnamefont {Speiser}},\ and\ \bibinfo {author} {\bibfnamefont {N.~P.}\ \bibnamefont {Stern}},\ }\bibfield  {title} {\bibinfo {title} {Electrical control of circular photogalvanic spin-valley photocurrent in a monolayer semiconductor},\ }\href@noop {} {\bibfield  {journal} {\bibinfo  {journal} {ACS Appl. Mater. Interfaces}\ }\textbf {\bibinfo {volume} {11}},\ \bibinfo {pages} {3334} (\bibinfo {year} {2018})}\BibitemShut {NoStop}%
\bibitem [{\citenamefont {Fregoso}(2019)}]{fregoso2019bulk}%
  \BibitemOpen
  \bibfield  {author} {\bibinfo {author} {\bibfnamefont {B.~M.}\ \bibnamefont {Fregoso}},\ }\bibfield  {title} {\bibinfo {title} {Bulk photovoltaic effects in the presence of a static electric field},\ }\href@noop {} {\bibfield  {journal} {\bibinfo  {journal} {Phys. Rev. B}\ }\textbf {\bibinfo {volume} {100}},\ \bibinfo {pages} {064301} (\bibinfo {year} {2019})}\BibitemShut {NoStop}%
\bibitem [{\citenamefont {Edelstein}(1990)}]{edelstein1990spin}%
  \BibitemOpen
  \bibfield  {author} {\bibinfo {author} {\bibfnamefont {V.~M.}\ \bibnamefont {Edelstein}},\ }\bibfield  {title} {\bibinfo {title} {Spin polarization of conduction electrons induced by electric current in two-dimensional asymmetric electron systems},\ }\href@noop {} {\bibfield  {journal} {\bibinfo  {journal} {Solid State Commun.}\ }\textbf {\bibinfo {volume} {73}},\ \bibinfo {pages} {233} (\bibinfo {year} {1990})}\BibitemShut {NoStop}%
\bibitem [{\citenamefont {Rojas-S{\'a}nchez}\ \emph {et~al.}(2016)\citenamefont {Rojas-S{\'a}nchez}, \citenamefont {Oyarz{\'u}n}, \citenamefont {Fu}, \citenamefont {Marty}, \citenamefont {Vergnaud}, \citenamefont {Gambarelli}, \citenamefont {Vila}, \citenamefont {Jamet}, \citenamefont {Ohtsubo}, \citenamefont {Taleb-Ibrahimi} \emph {et~al.}}]{rojas2016spin}%
  \BibitemOpen
  \bibfield  {author} {\bibinfo {author} {\bibfnamefont {J.-C.}\ \bibnamefont {Rojas-S{\'a}nchez}}, \bibinfo {author} {\bibfnamefont {S.}~\bibnamefont {Oyarz{\'u}n}}, \bibinfo {author} {\bibfnamefont {Y.}~\bibnamefont {Fu}}, \bibinfo {author} {\bibfnamefont {A.}~\bibnamefont {Marty}}, \bibinfo {author} {\bibfnamefont {C.}~\bibnamefont {Vergnaud}}, \bibinfo {author} {\bibfnamefont {S.}~\bibnamefont {Gambarelli}}, \bibinfo {author} {\bibfnamefont {L.}~\bibnamefont {Vila}}, \bibinfo {author} {\bibfnamefont {M.}~\bibnamefont {Jamet}}, \bibinfo {author} {\bibfnamefont {Y.}~\bibnamefont {Ohtsubo}}, \bibinfo {author} {\bibfnamefont {A.}~\bibnamefont {Taleb-Ibrahimi}}, \emph {et~al.},\ }\bibfield  {title} {\bibinfo {title} {Spin to charge conversion at room temperature by spin pumping into a new type of topological insulator: $\alpha$-sn films},\ }\href@noop {} {\bibfield  {journal} {\bibinfo  {journal} {Phys. Rev. Lett.}\ }\textbf {\bibinfo {volume} {116}},\ \bibinfo {pages} {096602} (\bibinfo {year}
  {2016})}\BibitemShut {NoStop}%
\bibitem [{\citenamefont {Shao}\ \emph {et~al.}(2016)\citenamefont {Shao}, \citenamefont {Yu}, \citenamefont {Lan}, \citenamefont {Shi}, \citenamefont {Li}, \citenamefont {Zheng}, \citenamefont {Zhu}, \citenamefont {Li}, \citenamefont {Amiri},\ and\ \citenamefont {Wang}}]{shao2016strong}%
  \BibitemOpen
  \bibfield  {author} {\bibinfo {author} {\bibfnamefont {Q.}~\bibnamefont {Shao}}, \bibinfo {author} {\bibfnamefont {G.}~\bibnamefont {Yu}}, \bibinfo {author} {\bibfnamefont {Y.-W.}\ \bibnamefont {Lan}}, \bibinfo {author} {\bibfnamefont {Y.}~\bibnamefont {Shi}}, \bibinfo {author} {\bibfnamefont {M.-Y.}\ \bibnamefont {Li}}, \bibinfo {author} {\bibfnamefont {C.}~\bibnamefont {Zheng}}, \bibinfo {author} {\bibfnamefont {X.}~\bibnamefont {Zhu}}, \bibinfo {author} {\bibfnamefont {L.-J.}\ \bibnamefont {Li}}, \bibinfo {author} {\bibfnamefont {P.~K.}\ \bibnamefont {Amiri}},\ and\ \bibinfo {author} {\bibfnamefont {K.~L.}\ \bibnamefont {Wang}},\ }\bibfield  {title} {\bibinfo {title} {Strong rashba-edelstein effect-induced spin-orbit torques in monolayer transition metal dichalcogenide/ferromagnet bilayers},\ }\href@noop {} {\bibfield  {journal} {\bibinfo  {journal} {Nano Lett.}\ }\textbf {\bibinfo {volume} {16}},\ \bibinfo {pages} {7514} (\bibinfo {year} {2016})}\BibitemShut {NoStop}%
\bibitem [{\citenamefont {Sun}\ \emph {et~al.}(2019)\citenamefont {Sun}, \citenamefont {Zhang}, \citenamefont {Kavand}, \citenamefont {Wang}, \citenamefont {Malissa}, \citenamefont {Liu}, \citenamefont {Popli}, \citenamefont {Singh}, \citenamefont {Vardeny}, \citenamefont {Zhang} \emph {et~al.}}]{sun2019surface}%
  \BibitemOpen
  \bibfield  {author} {\bibinfo {author} {\bibfnamefont {D.}~\bibnamefont {Sun}}, \bibinfo {author} {\bibfnamefont {C.}~\bibnamefont {Zhang}}, \bibinfo {author} {\bibfnamefont {M.}~\bibnamefont {Kavand}}, \bibinfo {author} {\bibfnamefont {J.}~\bibnamefont {Wang}}, \bibinfo {author} {\bibfnamefont {H.}~\bibnamefont {Malissa}}, \bibinfo {author} {\bibfnamefont {H.}~\bibnamefont {Liu}}, \bibinfo {author} {\bibfnamefont {H.}~\bibnamefont {Popli}}, \bibinfo {author} {\bibfnamefont {J.}~\bibnamefont {Singh}}, \bibinfo {author} {\bibfnamefont {S.~R.}\ \bibnamefont {Vardeny}}, \bibinfo {author} {\bibfnamefont {W.}~\bibnamefont {Zhang}}, \emph {et~al.},\ }\bibfield  {title} {\bibinfo {title} {{Surface-enhanced spin current to charge current conversion efficiency in CH3NH3PbBr3-based devices}},\ }\href@noop {} {\bibfield  {journal} {\bibinfo  {journal} {J. Chem. Phys.}\ }\textbf {\bibinfo {volume} {151}},\ \bibinfo {pages} {174709} (\bibinfo {year} {2019})}\BibitemShut {NoStop}%
\bibitem [{\citenamefont {Trier}\ \emph {et~al.}(2019)\citenamefont {Trier}, \citenamefont {Vaz}, \citenamefont {Bruneel}, \citenamefont {No{\"e}l}, \citenamefont {Fert}, \citenamefont {Vila}, \citenamefont {Attan{\'e}}, \citenamefont {Barth{\'e}l{\'e}my}, \citenamefont {Gabay}, \citenamefont {Jaffres} \emph {et~al.}}]{trier2019electric}%
  \BibitemOpen
  \bibfield  {author} {\bibinfo {author} {\bibfnamefont {F.}~\bibnamefont {Trier}}, \bibinfo {author} {\bibfnamefont {D.~C.}\ \bibnamefont {Vaz}}, \bibinfo {author} {\bibfnamefont {P.}~\bibnamefont {Bruneel}}, \bibinfo {author} {\bibfnamefont {P.}~\bibnamefont {No{\"e}l}}, \bibinfo {author} {\bibfnamefont {A.}~\bibnamefont {Fert}}, \bibinfo {author} {\bibfnamefont {L.}~\bibnamefont {Vila}}, \bibinfo {author} {\bibfnamefont {J.-P.}\ \bibnamefont {Attan{\'e}}}, \bibinfo {author} {\bibfnamefont {A.}~\bibnamefont {Barth{\'e}l{\'e}my}}, \bibinfo {author} {\bibfnamefont {M.}~\bibnamefont {Gabay}}, \bibinfo {author} {\bibfnamefont {H.}~\bibnamefont {Jaffres}}, \emph {et~al.},\ }\bibfield  {title} {\bibinfo {title} {{Electric-field control of spin current generation and detection in ferromagnet-free SrTiO3-based nanodevices}},\ }\href@noop {} {\bibfield  {journal} {\bibinfo  {journal} {Nano Lett.}\ }\textbf {\bibinfo {volume} {20}},\ \bibinfo {pages} {395} (\bibinfo {year} {2019})}\BibitemShut {NoStop}%
\bibitem [{\citenamefont {Chen}\ \emph {et~al.}(2022)\citenamefont {Chen}, \citenamefont {Ye}, \citenamefont {Zou}, \citenamefont {Gu}, \citenamefont {Xu},\ and\ \citenamefont {Duan}}]{chen_basic_2022}%
  \BibitemOpen
  \bibfield  {author} {\bibinfo {author} {\bibfnamefont {H.}~\bibnamefont {Chen}}, \bibinfo {author} {\bibfnamefont {M.}~\bibnamefont {Ye}}, \bibinfo {author} {\bibfnamefont {N.}~\bibnamefont {Zou}}, \bibinfo {author} {\bibfnamefont {B.-L.}\ \bibnamefont {Gu}}, \bibinfo {author} {\bibfnamefont {Y.}~\bibnamefont {Xu}},\ and\ \bibinfo {author} {\bibfnamefont {W.}~\bibnamefont {Duan}},\ }\bibfield  {title} {\bibinfo {title} {Basic formulation and first-principles implementation of nonlinear magneto-optical effects},\ }\href {https://doi.org/10.1103/PhysRevB.105.075123} {\bibfield  {journal} {\bibinfo  {journal} {Phys. Rev. B}\ }\textbf {\bibinfo {volume} {105}},\ \bibinfo {pages} {075123} (\bibinfo {year} {2022})}\BibitemShut {NoStop}%
\bibitem [{\citenamefont {Wang}\ and\ \citenamefont {Qian}(2020)}]{wang_electrically_2020}%
  \BibitemOpen
  \bibfield  {author} {\bibinfo {author} {\bibfnamefont {H.}~\bibnamefont {Wang}}\ and\ \bibinfo {author} {\bibfnamefont {X.}~\bibnamefont {Qian}},\ }\bibfield  {title} {\bibinfo {title} {{Electrically and magnetically switchable nonlinear photocurrent in PT-symmetric magnetic topological quantum materials}},\ }\href {https://doi.org/10.1038/s41524-020-00462-9} {\bibfield  {journal} {\bibinfo  {journal} {Npj Comput. Mater.}\ }\textbf {\bibinfo {volume} {6}},\ \bibinfo {pages} {1} (\bibinfo {year} {2020})}\BibitemShut {NoStop}%
\bibitem [{\citenamefont {Watanabe}\ and\ \citenamefont {Yanase}(2021)}]{watanabe_photocurrent_2021}%
  \BibitemOpen
  \bibfield  {author} {\bibinfo {author} {\bibfnamefont {H.}~\bibnamefont {Watanabe}}\ and\ \bibinfo {author} {\bibfnamefont {Y.}~\bibnamefont {Yanase}},\ }\bibfield  {title} {\bibinfo {title} {Photocurrent response in parity-time symmetric current-ordered states},\ }\href {https://doi.org/10.1103/PhysRevB.104.024416} {\bibfield  {journal} {\bibinfo  {journal} {Phys. Rev. B}\ }\textbf {\bibinfo {volume} {104}},\ \bibinfo {pages} {024416} (\bibinfo {year} {2021})}\BibitemShut {NoStop}%
\bibitem [{\citenamefont {Pi}\ \emph {et~al.}(2023)\citenamefont {Pi}, \citenamefont {Zhang},\ and\ \citenamefont {Weng}}]{pi2023magnetic}%
  \BibitemOpen
  \bibfield  {author} {\bibinfo {author} {\bibfnamefont {H.}~\bibnamefont {Pi}}, \bibinfo {author} {\bibfnamefont {S.}~\bibnamefont {Zhang}},\ and\ \bibinfo {author} {\bibfnamefont {H.}~\bibnamefont {Weng}},\ }\bibfield  {title} {\bibinfo {title} {{Magnetic bulk photovoltaic effect as a probe of magnetic structures of EuSn2As2}},\ }\href@noop {} {\bibfield  {journal} {\bibinfo  {journal} {Quantum Front.}\ }\textbf {\bibinfo {volume} {2}},\ \bibinfo {pages} {1} (\bibinfo {year} {2023})}\BibitemShut {NoStop}%
\bibitem [{\citenamefont {Zhang}\ \emph {et~al.}(2019)\citenamefont {Zhang}, \citenamefont {Holder}, \citenamefont {Ishizuka}, \citenamefont {de~Juan}, \citenamefont {Nagaosa}, \citenamefont {Felser},\ and\ \citenamefont {Yan}}]{zhang_switchable_2019}%
  \BibitemOpen
  \bibfield  {author} {\bibinfo {author} {\bibfnamefont {Y.}~\bibnamefont {Zhang}}, \bibinfo {author} {\bibfnamefont {T.}~\bibnamefont {Holder}}, \bibinfo {author} {\bibfnamefont {H.}~\bibnamefont {Ishizuka}}, \bibinfo {author} {\bibfnamefont {F.}~\bibnamefont {de~Juan}}, \bibinfo {author} {\bibfnamefont {N.}~\bibnamefont {Nagaosa}}, \bibinfo {author} {\bibfnamefont {C.}~\bibnamefont {Felser}},\ and\ \bibinfo {author} {\bibfnamefont {B.}~\bibnamefont {Yan}},\ }\bibfield  {title} {\bibinfo {title} {Switchable magnetic bulk photovoltaic effect in the two-dimensional magnet {CrI3}},\ }\href {https://doi.org/10.1038/s41467-019-11832-3} {\bibfield  {journal} {\bibinfo  {journal} {Nature Commun.}\ }\textbf {\bibinfo {volume} {10}},\ \bibinfo {pages} {3783} (\bibinfo {year} {2019})}\BibitemShut {NoStop}%
\bibitem [{\citenamefont {Fei}\ \emph {et~al.}(2020)\citenamefont {Fei}, \citenamefont {Song},\ and\ \citenamefont {Yang}}]{fei_giant_2020}%
  \BibitemOpen
  \bibfield  {author} {\bibinfo {author} {\bibfnamefont {R.}~\bibnamefont {Fei}}, \bibinfo {author} {\bibfnamefont {W.}~\bibnamefont {Song}},\ and\ \bibinfo {author} {\bibfnamefont {L.}~\bibnamefont {Yang}},\ }\bibfield  {title} {\bibinfo {title} {Giant photogalvanic effect and second-harmonic generation in magnetic axion insulators},\ }\href {https://doi.org/10.1103/PhysRevB.102.035440} {\bibfield  {journal} {\bibinfo  {journal} {Phys. Rev. B}\ }\textbf {\bibinfo {volume} {102}},\ \bibinfo {pages} {035440} (\bibinfo {year} {2020})}\BibitemShut {NoStop}%
\bibitem [{\citenamefont {Fei}\ \emph {et~al.}(2021)\citenamefont {Fei}, \citenamefont {Song}, \citenamefont {Pusey-Nazzaro},\ and\ \citenamefont {Yang}}]{fei_pt-symmetry-enabled_2021}%
  \BibitemOpen
  \bibfield  {author} {\bibinfo {author} {\bibfnamefont {R.}~\bibnamefont {Fei}}, \bibinfo {author} {\bibfnamefont {W.}~\bibnamefont {Song}}, \bibinfo {author} {\bibfnamefont {L.}~\bibnamefont {Pusey-Nazzaro}},\ and\ \bibinfo {author} {\bibfnamefont {L.}~\bibnamefont {Yang}},\ }\bibfield  {title} {\bibinfo {title} {{PT}-{Symmetry}-{Enabled} {Spin} {Circular} {Photogalvanic} {Effect} in {Antiferromagnetic} {Insulators}},\ }\href {https://doi.org/10.1103/PhysRevLett.127.207402} {\bibfield  {journal} {\bibinfo  {journal} {Phys. Rev. Lett.}\ }\textbf {\bibinfo {volume} {127}},\ \bibinfo {pages} {207402} (\bibinfo {year} {2021})}\BibitemShut {NoStop}%
\bibitem [{\citenamefont {Xue}\ \emph {et~al.}(2023)\citenamefont {Xue}, \citenamefont {Mu}, \citenamefont {Sun},\ and\ \citenamefont {Zhou}}]{xue_valley_2023}%
  \BibitemOpen
  \bibfield  {author} {\bibinfo {author} {\bibfnamefont {Q.}~\bibnamefont {Xue}}, \bibinfo {author} {\bibfnamefont {X.}~\bibnamefont {Mu}}, \bibinfo {author} {\bibfnamefont {Y.}~\bibnamefont {Sun}},\ and\ \bibinfo {author} {\bibfnamefont {J.}~\bibnamefont {Zhou}},\ }\bibfield  {title} {\bibinfo {title} {{Valley contrasting bulk photovoltaic effect in a PT-symmetric MnPSe3 monolayer}},\ }\href {https://doi.org/10.1103/PhysRevB.107.245404} {\bibfield  {journal} {\bibinfo  {journal} {Phys. Rev. B}\ }\textbf {\bibinfo {volume} {107}},\ \bibinfo {pages} {245404} (\bibinfo {year} {2023})}\BibitemShut {NoStop}%
\bibitem [{\citenamefont {Liu}\ \emph {et~al.}(2023)\citenamefont {Liu}, \citenamefont {Liu}, \citenamefont {Cheng}, \citenamefont {Cui},\ and\ \citenamefont {Hu}}]{liu_switchable_2023}%
  \BibitemOpen
  \bibfield  {author} {\bibinfo {author} {\bibfnamefont {L.}~\bibnamefont {Liu}}, \bibinfo {author} {\bibfnamefont {W.}~\bibnamefont {Liu}}, \bibinfo {author} {\bibfnamefont {B.}~\bibnamefont {Cheng}}, \bibinfo {author} {\bibfnamefont {B.}~\bibnamefont {Cui}},\ and\ \bibinfo {author} {\bibfnamefont {J.}~\bibnamefont {Hu}},\ }\bibfield  {title} {\bibinfo {title} {{Switchable Giant Bulk Photocurrents and Photo-spin-currents in Monolayer PT-Symmetric Antiferromagnet MnPSe3}},\ }\href {https://doi.org/10.1021/acs.jpclett.2c03383} {\bibfield  {journal} {\bibinfo  {journal} {J. Phys. Chem. Lett.}\ }\textbf {\bibinfo {volume} {14}},\ \bibinfo {pages} {370} (\bibinfo {year} {2023})}\BibitemShut {NoStop}%
\bibitem [{\citenamefont {Fu}\ \emph {et~al.}(2011)\citenamefont {Fu}, \citenamefont {Zou}, \citenamefont {Wang}, \citenamefont {Zhang}, \citenamefont {Yu},\ and\ \citenamefont {Wu}}]{fu2011electrothermal}%
  \BibitemOpen
  \bibfield  {author} {\bibinfo {author} {\bibfnamefont {D.}~\bibnamefont {Fu}}, \bibinfo {author} {\bibfnamefont {J.}~\bibnamefont {Zou}}, \bibinfo {author} {\bibfnamefont {K.}~\bibnamefont {Wang}}, \bibinfo {author} {\bibfnamefont {R.}~\bibnamefont {Zhang}}, \bibinfo {author} {\bibfnamefont {D.}~\bibnamefont {Yu}},\ and\ \bibinfo {author} {\bibfnamefont {J.}~\bibnamefont {Wu}},\ }\bibfield  {title} {\bibinfo {title} {Electrothermal dynamics of semiconductor nanowires under local carrier modulation},\ }\href@noop {} {\bibfield  {journal} {\bibinfo  {journal} {Nano Lett.}\ }\textbf {\bibinfo {volume} {11}},\ \bibinfo {pages} {3809} (\bibinfo {year} {2011})}\BibitemShut {NoStop}%
\bibitem [{\citenamefont {Graham}\ and\ \citenamefont {Yu}(2013)}]{graham2013scanning}%
  \BibitemOpen
  \bibfield  {author} {\bibinfo {author} {\bibfnamefont {R.}~\bibnamefont {Graham}}\ and\ \bibinfo {author} {\bibfnamefont {D.}~\bibnamefont {Yu}},\ }\bibfield  {title} {\bibinfo {title} {Scanning photocurrent microscopy in semiconductor nanostructures},\ }\href@noop {} {\bibfield  {journal} {\bibinfo  {journal} {Mod. Phys. Lett. B}\ }\textbf {\bibinfo {volume} {27}},\ \bibinfo {pages} {1330018} (\bibinfo {year} {2013})}\BibitemShut {NoStop}%
\bibitem [{\citenamefont {Xiao}\ \emph {et~al.}(2016)\citenamefont {Xiao}, \citenamefont {Hou}, \citenamefont {Fu}, \citenamefont {Peng}, \citenamefont {Wang}, \citenamefont {Gonzalez}, \citenamefont {Jin},\ and\ \citenamefont {Yu}}]{xiao2016photocurrent}%
  \BibitemOpen
  \bibfield  {author} {\bibinfo {author} {\bibfnamefont {R.}~\bibnamefont {Xiao}}, \bibinfo {author} {\bibfnamefont {Y.}~\bibnamefont {Hou}}, \bibinfo {author} {\bibfnamefont {Y.}~\bibnamefont {Fu}}, \bibinfo {author} {\bibfnamefont {X.}~\bibnamefont {Peng}}, \bibinfo {author} {\bibfnamefont {Q.}~\bibnamefont {Wang}}, \bibinfo {author} {\bibfnamefont {E.}~\bibnamefont {Gonzalez}}, \bibinfo {author} {\bibfnamefont {S.}~\bibnamefont {Jin}},\ and\ \bibinfo {author} {\bibfnamefont {D.}~\bibnamefont {Yu}},\ }\bibfield  {title} {\bibinfo {title} {Photocurrent mapping in single-crystal methylammonium lead iodide perovskite nanostructures},\ }\href {https://doi.org/10.1021/acs.nanolett.6b03782} {\bibfield  {journal} {\bibinfo  {journal} {Nano Lett.}\ }\textbf {\bibinfo {volume} {16}},\ \bibinfo {pages} {7710} (\bibinfo {year} {2016})}\BibitemShut {NoStop}%
\bibitem [{\citenamefont {Wang}\ \emph {et~al.}(2017)\citenamefont {Wang}, \citenamefont {Ling}, \citenamefont {Chiu}, \citenamefont {Du}, \citenamefont {Barreda}, \citenamefont {Perez-Orive}, \citenamefont {Ma}, \citenamefont {Xiong},\ and\ \citenamefont {Gao}}]{wang2017dynamic}%
  \BibitemOpen
  \bibfield  {author} {\bibinfo {author} {\bibfnamefont {X.}~\bibnamefont {Wang}}, \bibinfo {author} {\bibfnamefont {Y.}~\bibnamefont {Ling}}, \bibinfo {author} {\bibfnamefont {Y.-C.}\ \bibnamefont {Chiu}}, \bibinfo {author} {\bibfnamefont {Y.}~\bibnamefont {Du}}, \bibinfo {author} {\bibfnamefont {J.~L.}\ \bibnamefont {Barreda}}, \bibinfo {author} {\bibfnamefont {F.}~\bibnamefont {Perez-Orive}}, \bibinfo {author} {\bibfnamefont {B.}~\bibnamefont {Ma}}, \bibinfo {author} {\bibfnamefont {P.}~\bibnamefont {Xiong}},\ and\ \bibinfo {author} {\bibfnamefont {H.}~\bibnamefont {Gao}},\ }\bibfield  {title} {\bibinfo {title} {Dynamic electronic junctions in organic--inorganic hybrid perovskites},\ }\href@noop {} {\bibfield  {journal} {\bibinfo  {journal} {Nano Lett.}\ }\textbf {\bibinfo {volume} {17}},\ \bibinfo {pages} {4831} (\bibinfo {year} {2017})}\BibitemShut {NoStop}%
\bibitem [{\citenamefont {Wang}\ \emph {et~al.}(2023)\citenamefont {Wang}, \citenamefont {Zhu}, \citenamefont {Travaglini}, \citenamefont {Sun}, \citenamefont {Savrasov}, \citenamefont {Hahn}, \citenamefont {van Benthem},\ and\ \citenamefont {Yu}}]{wang2023spatially}%
  \BibitemOpen
  \bibfield  {author} {\bibinfo {author} {\bibfnamefont {B.~M.}\ \bibnamefont {Wang}}, \bibinfo {author} {\bibfnamefont {Y.}~\bibnamefont {Zhu}}, \bibinfo {author} {\bibfnamefont {H.~C.}\ \bibnamefont {Travaglini}}, \bibinfo {author} {\bibfnamefont {R.}~\bibnamefont {Sun}}, \bibinfo {author} {\bibfnamefont {S.~Y.}\ \bibnamefont {Savrasov}}, \bibinfo {author} {\bibfnamefont {W.}~\bibnamefont {Hahn}}, \bibinfo {author} {\bibfnamefont {K.}~\bibnamefont {van Benthem}},\ and\ \bibinfo {author} {\bibfnamefont {D.}~\bibnamefont {Yu}},\ }\bibfield  {title} {\bibinfo {title} {{Spatially dispersive helicity-dependent photocurrent in Dirac semimetal Cd3As2 nanobelts}},\ }\href@noop {} {\bibfield  {journal} {\bibinfo  {journal} {Phys. Rev. B}\ }\textbf {\bibinfo {volume} {108}},\ \bibinfo {pages} {165405} (\bibinfo {year} {2023})}\BibitemShut {NoStop}%
\bibitem [{\citenamefont {McClintock}\ \emph {et~al.}(2020)\citenamefont {McClintock}, \citenamefont {Xiao}, \citenamefont {Hou}, \citenamefont {Gibson}, \citenamefont {Travaglini}, \citenamefont {Abramovitch}, \citenamefont {Tan}, \citenamefont {Senger}, \citenamefont {Fu}, \citenamefont {Jin} \emph {et~al.}}]{mcclintock2020temperature}%
  \BibitemOpen
  \bibfield  {author} {\bibinfo {author} {\bibfnamefont {L.}~\bibnamefont {McClintock}}, \bibinfo {author} {\bibfnamefont {R.}~\bibnamefont {Xiao}}, \bibinfo {author} {\bibfnamefont {Y.}~\bibnamefont {Hou}}, \bibinfo {author} {\bibfnamefont {C.}~\bibnamefont {Gibson}}, \bibinfo {author} {\bibfnamefont {H.~C.}\ \bibnamefont {Travaglini}}, \bibinfo {author} {\bibfnamefont {D.}~\bibnamefont {Abramovitch}}, \bibinfo {author} {\bibfnamefont {L.~Z.}\ \bibnamefont {Tan}}, \bibinfo {author} {\bibfnamefont {R.~T.}\ \bibnamefont {Senger}}, \bibinfo {author} {\bibfnamefont {Y.}~\bibnamefont {Fu}}, \bibinfo {author} {\bibfnamefont {S.}~\bibnamefont {Jin}}, \emph {et~al.},\ }\bibfield  {title} {\bibinfo {title} {Temperature and gate dependence of carrier diffusion in single crystal methylammonium lead iodide perovskite microstructures},\ }\href@noop {} {\bibfield  {journal} {\bibinfo  {journal} {J. Phys. Chem. Lett.}\ }\textbf {\bibinfo {volume} {11}},\ \bibinfo {pages} {1000} (\bibinfo {year} {2020})}\BibitemShut
  {NoStop}%
\bibitem [{\citenamefont {Giovanni}\ \emph {et~al.}(2015)\citenamefont {Giovanni}, \citenamefont {Ma}, \citenamefont {Chua}, \citenamefont {Gratzel}, \citenamefont {Ramesh}, \citenamefont {Mhaisalkar}, \citenamefont {Mathews},\ and\ \citenamefont {Sum}}]{giovanni2015highly}%
  \BibitemOpen
  \bibfield  {author} {\bibinfo {author} {\bibfnamefont {D.}~\bibnamefont {Giovanni}}, \bibinfo {author} {\bibfnamefont {H.}~\bibnamefont {Ma}}, \bibinfo {author} {\bibfnamefont {J.}~\bibnamefont {Chua}}, \bibinfo {author} {\bibfnamefont {M.}~\bibnamefont {Gratzel}}, \bibinfo {author} {\bibfnamefont {R.}~\bibnamefont {Ramesh}}, \bibinfo {author} {\bibfnamefont {S.}~\bibnamefont {Mhaisalkar}}, \bibinfo {author} {\bibfnamefont {N.}~\bibnamefont {Mathews}},\ and\ \bibinfo {author} {\bibfnamefont {T.~C.}\ \bibnamefont {Sum}},\ }\bibfield  {title} {\bibinfo {title} {{Highly spin-polarized carrier dynamics and ultralarge photoinduced magnetization in CH3NH3PbI3 perovskite thin films}},\ }\href@noop {} {\bibfield  {journal} {\bibinfo  {journal} {Nano Lett.}\ }\textbf {\bibinfo {volume} {15}},\ \bibinfo {pages} {1553} (\bibinfo {year} {2015})}\BibitemShut {NoStop}%
\bibitem [{\citenamefont {Odenthal}\ \emph {et~al.}(2017)\citenamefont {Odenthal}, \citenamefont {Talmadge}, \citenamefont {Gundlach}, \citenamefont {Wang}, \citenamefont {Zhang}, \citenamefont {Sun}, \citenamefont {Yu}, \citenamefont {Valy~Vardeny},\ and\ \citenamefont {Li}}]{odenthal2017spin}%
  \BibitemOpen
  \bibfield  {author} {\bibinfo {author} {\bibfnamefont {P.}~\bibnamefont {Odenthal}}, \bibinfo {author} {\bibfnamefont {W.}~\bibnamefont {Talmadge}}, \bibinfo {author} {\bibfnamefont {N.}~\bibnamefont {Gundlach}}, \bibinfo {author} {\bibfnamefont {R.}~\bibnamefont {Wang}}, \bibinfo {author} {\bibfnamefont {C.}~\bibnamefont {Zhang}}, \bibinfo {author} {\bibfnamefont {D.}~\bibnamefont {Sun}}, \bibinfo {author} {\bibfnamefont {Z.-G.}\ \bibnamefont {Yu}}, \bibinfo {author} {\bibfnamefont {Z.}~\bibnamefont {Valy~Vardeny}},\ and\ \bibinfo {author} {\bibfnamefont {Y.~S.}\ \bibnamefont {Li}},\ }\bibfield  {title} {\bibinfo {title} {Spin-polarized exciton quantum beating in hybrid organic--inorganic perovskites},\ }\href@noop {} {\bibfield  {journal} {\bibinfo  {journal} {Nat. Phys.}\ }\textbf {\bibinfo {volume} {13}},\ \bibinfo {pages} {894} (\bibinfo {year} {2017})}\BibitemShut {NoStop}%
\bibitem [{\citenamefont {Kirstein}\ \emph {et~al.}(2022)\citenamefont {Kirstein}, \citenamefont {Yakovlev}, \citenamefont {Zhukov}, \citenamefont {Hocker}, \citenamefont {Dyakonov},\ and\ \citenamefont {Bayer}}]{kirstein2022spin}%
  \BibitemOpen
  \bibfield  {author} {\bibinfo {author} {\bibfnamefont {E.}~\bibnamefont {Kirstein}}, \bibinfo {author} {\bibfnamefont {D.~R.}\ \bibnamefont {Yakovlev}}, \bibinfo {author} {\bibfnamefont {E.~A.}\ \bibnamefont {Zhukov}}, \bibinfo {author} {\bibfnamefont {J.}~\bibnamefont {Hocker}}, \bibinfo {author} {\bibfnamefont {V.}~\bibnamefont {Dyakonov}},\ and\ \bibinfo {author} {\bibfnamefont {M.}~\bibnamefont {Bayer}},\ }\bibfield  {title} {\bibinfo {title} {Spin dynamics of electrons and holes interacting with nuclei in mapbi3 perovskite single crystals},\ }\href@noop {} {\bibfield  {journal} {\bibinfo  {journal} {ACS Photonics}\ }\textbf {\bibinfo {volume} {9}},\ \bibinfo {pages} {1375} (\bibinfo {year} {2022})}\BibitemShut {NoStop}%
\bibitem [{\citenamefont {Xu}\ \emph {et~al.}(2024)\citenamefont {Xu}, \citenamefont {Li}, \citenamefont {Huynh}, \citenamefont {Fadel}, \citenamefont {Huang}, \citenamefont {Sundararaman}, \citenamefont {Vardeny},\ and\ \citenamefont {Ping}}]{xu2024spin}%
  \BibitemOpen
  \bibfield  {author} {\bibinfo {author} {\bibfnamefont {J.}~\bibnamefont {Xu}}, \bibinfo {author} {\bibfnamefont {K.}~\bibnamefont {Li}}, \bibinfo {author} {\bibfnamefont {U.~N.}\ \bibnamefont {Huynh}}, \bibinfo {author} {\bibfnamefont {M.}~\bibnamefont {Fadel}}, \bibinfo {author} {\bibfnamefont {J.}~\bibnamefont {Huang}}, \bibinfo {author} {\bibfnamefont {R.}~\bibnamefont {Sundararaman}}, \bibinfo {author} {\bibfnamefont {V.}~\bibnamefont {Vardeny}},\ and\ \bibinfo {author} {\bibfnamefont {Y.}~\bibnamefont {Ping}},\ }\bibfield  {title} {\bibinfo {title} {How spin relaxes and dephases in bulk halide perovskites},\ }\href@noop {} {\bibfield  {journal} {\bibinfo  {journal} {Nature Commun.}\ }\textbf {\bibinfo {volume} {15}},\ \bibinfo {pages} {188} (\bibinfo {year} {2024})}\BibitemShut {NoStop}%
\bibitem [{\citenamefont {Yang}\ \emph {et~al.}(2019)\citenamefont {Yang}, \citenamefont {Feng}, \citenamefont {Li}, \citenamefont {Li}, \citenamefont {Xiong}, \citenamefont {Cao},\ and\ \citenamefont {Gao}}]{yang2019unexpected}%
  \BibitemOpen
  \bibfield  {author} {\bibinfo {author} {\bibfnamefont {Y.}~\bibnamefont {Yang}}, \bibinfo {author} {\bibfnamefont {S.}~\bibnamefont {Feng}}, \bibinfo {author} {\bibfnamefont {Z.}~\bibnamefont {Li}}, \bibinfo {author} {\bibfnamefont {T.}~\bibnamefont {Li}}, \bibinfo {author} {\bibfnamefont {Y.}~\bibnamefont {Xiong}}, \bibinfo {author} {\bibfnamefont {L.}~\bibnamefont {Cao}},\ and\ \bibinfo {author} {\bibfnamefont {X.}~\bibnamefont {Gao}},\ }\bibfield  {title} {\bibinfo {title} {Unexpected outstanding room temperature spin transport verified in organic--inorganic hybrid perovskite film},\ }\href@noop {} {\bibfield  {journal} {\bibinfo  {journal} {J. Phys. Chem. Lett.}\ }\textbf {\bibinfo {volume} {10}},\ \bibinfo {pages} {4422} (\bibinfo {year} {2019})}\BibitemShut {NoStop}%
\bibitem [{\citenamefont {Frost}(2017)}]{frost2017calculating}%
  \BibitemOpen
  \bibfield  {author} {\bibinfo {author} {\bibfnamefont {J.~M.}\ \bibnamefont {Frost}},\ }\bibfield  {title} {\bibinfo {title} {Calculating polaron mobility in halide perovskites},\ }\href@noop {} {\bibfield  {journal} {\bibinfo  {journal} {Phys. Rev. B}\ }\textbf {\bibinfo {volume} {96}},\ \bibinfo {pages} {195202} (\bibinfo {year} {2017})}\BibitemShut {NoStop}%
\bibitem [{\citenamefont {Tang}\ \emph {et~al.}(2021)\citenamefont {Tang}, \citenamefont {Li}, \citenamefont {Weeden}, \citenamefont {Song}, \citenamefont {McClintock}, \citenamefont {Xiao}, \citenamefont {Senger},\ and\ \citenamefont {Yu}}]{tang2021transport}%
  \BibitemOpen
  \bibfield  {author} {\bibinfo {author} {\bibfnamefont {K.~W.}\ \bibnamefont {Tang}}, \bibinfo {author} {\bibfnamefont {S.}~\bibnamefont {Li}}, \bibinfo {author} {\bibfnamefont {S.}~\bibnamefont {Weeden}}, \bibinfo {author} {\bibfnamefont {Z.}~\bibnamefont {Song}}, \bibinfo {author} {\bibfnamefont {L.}~\bibnamefont {McClintock}}, \bibinfo {author} {\bibfnamefont {R.}~\bibnamefont {Xiao}}, \bibinfo {author} {\bibfnamefont {R.~T.}\ \bibnamefont {Senger}},\ and\ \bibinfo {author} {\bibfnamefont {D.}~\bibnamefont {Yu}},\ }\bibfield  {title} {\bibinfo {title} {Transport modeling of locally photogenerated excitons in halide perovskites},\ }\href@noop {} {\bibfield  {journal} {\bibinfo  {journal} {J. Phys. Chem. Lett.}\ }\textbf {\bibinfo {volume} {12}},\ \bibinfo {pages} {3951} (\bibinfo {year} {2021})}\BibitemShut {NoStop}%
\bibitem [{\citenamefont {McClintock}\ \emph {et~al.}(2022)\citenamefont {McClintock}, \citenamefont {Song}, \citenamefont {Travaglini}, \citenamefont {Senger}, \citenamefont {Chandrasekaran}, \citenamefont {Htoon}, \citenamefont {Yarotski},\ and\ \citenamefont {Yu}}]{mcclintock2022highly}%
  \BibitemOpen
  \bibfield  {author} {\bibinfo {author} {\bibfnamefont {L.}~\bibnamefont {McClintock}}, \bibinfo {author} {\bibfnamefont {Z.}~\bibnamefont {Song}}, \bibinfo {author} {\bibfnamefont {H.~C.}\ \bibnamefont {Travaglini}}, \bibinfo {author} {\bibfnamefont {R.~T.}\ \bibnamefont {Senger}}, \bibinfo {author} {\bibfnamefont {V.}~\bibnamefont {Chandrasekaran}}, \bibinfo {author} {\bibfnamefont {H.}~\bibnamefont {Htoon}}, \bibinfo {author} {\bibfnamefont {D.}~\bibnamefont {Yarotski}},\ and\ \bibinfo {author} {\bibfnamefont {D.}~\bibnamefont {Yu}},\ }\bibfield  {title} {\bibinfo {title} {Highly mobile excitons in single crystal methylammonium lead tribromide perovskite microribbons},\ }\href@noop {} {\bibfield  {journal} {\bibinfo  {journal} {J. Phys. Chem. Lett.}\ }\textbf {\bibinfo {volume} {13}},\ \bibinfo {pages} {3698} (\bibinfo {year} {2022})}\BibitemShut {NoStop}%
\bibitem [{\citenamefont {Frost}\ \emph {et~al.}(2014)\citenamefont {Frost}, \citenamefont {Butler}, \citenamefont {Brivio}, \citenamefont {Hendon}, \citenamefont {Van~Schilfgaarde},\ and\ \citenamefont {Walsh}}]{frost2014atomistic}%
  \BibitemOpen
  \bibfield  {author} {\bibinfo {author} {\bibfnamefont {J.~M.}\ \bibnamefont {Frost}}, \bibinfo {author} {\bibfnamefont {K.~T.}\ \bibnamefont {Butler}}, \bibinfo {author} {\bibfnamefont {F.}~\bibnamefont {Brivio}}, \bibinfo {author} {\bibfnamefont {C.~H.}\ \bibnamefont {Hendon}}, \bibinfo {author} {\bibfnamefont {M.}~\bibnamefont {Van~Schilfgaarde}},\ and\ \bibinfo {author} {\bibfnamefont {A.}~\bibnamefont {Walsh}},\ }\bibfield  {title} {\bibinfo {title} {Atomistic origins of high-performance in hybrid halide perovskite solar cells},\ }\href@noop {} {\bibfield  {journal} {\bibinfo  {journal} {Nano Lett.}\ }\textbf {\bibinfo {volume} {14}},\ \bibinfo {pages} {2584} (\bibinfo {year} {2014})}\BibitemShut {NoStop}%
\bibitem [{\citenamefont {Garten}\ \emph {et~al.}(2019)\citenamefont {Garten}, \citenamefont {Moore}, \citenamefont {Nanayakkara}, \citenamefont {Dwaraknath}, \citenamefont {Schulz}, \citenamefont {Wands}, \citenamefont {Rockett}, \citenamefont {Newell}, \citenamefont {Persson}, \citenamefont {Trolier-McKinstry} \emph {et~al.}}]{garten2019existence}%
  \BibitemOpen
  \bibfield  {author} {\bibinfo {author} {\bibfnamefont {L.~M.}\ \bibnamefont {Garten}}, \bibinfo {author} {\bibfnamefont {D.~T.}\ \bibnamefont {Moore}}, \bibinfo {author} {\bibfnamefont {S.~U.}\ \bibnamefont {Nanayakkara}}, \bibinfo {author} {\bibfnamefont {S.}~\bibnamefont {Dwaraknath}}, \bibinfo {author} {\bibfnamefont {P.}~\bibnamefont {Schulz}}, \bibinfo {author} {\bibfnamefont {J.}~\bibnamefont {Wands}}, \bibinfo {author} {\bibfnamefont {A.}~\bibnamefont {Rockett}}, \bibinfo {author} {\bibfnamefont {B.}~\bibnamefont {Newell}}, \bibinfo {author} {\bibfnamefont {K.~A.}\ \bibnamefont {Persson}}, \bibinfo {author} {\bibfnamefont {S.}~\bibnamefont {Trolier-McKinstry}}, \emph {et~al.},\ }\bibfield  {title} {\bibinfo {title} {The existence and impact of persistent ferroelectric domains in mapbi3},\ }\href@noop {} {\bibfield  {journal} {\bibinfo  {journal} {Sci. Adv.}\ }\textbf {\bibinfo {volume} {5}},\ \bibinfo {pages} {eaas9311} (\bibinfo {year} {2019})}\BibitemShut {NoStop}%
\bibitem [{\citenamefont {Kim}\ \emph {et~al.}(2014)\citenamefont {Kim}, \citenamefont {Im}, \citenamefont {Freeman}, \citenamefont {Ihm},\ and\ \citenamefont {Jin}}]{kim2014switchable}%
  \BibitemOpen
  \bibfield  {author} {\bibinfo {author} {\bibfnamefont {M.}~\bibnamefont {Kim}}, \bibinfo {author} {\bibfnamefont {J.}~\bibnamefont {Im}}, \bibinfo {author} {\bibfnamefont {A.~J.}\ \bibnamefont {Freeman}}, \bibinfo {author} {\bibfnamefont {J.}~\bibnamefont {Ihm}},\ and\ \bibinfo {author} {\bibfnamefont {H.}~\bibnamefont {Jin}},\ }\bibfield  {title} {\bibinfo {title} {Switchable s= 1/2 and j= 1/2 rashba bands in ferroelectric halide perovskites},\ }\href@noop {} {\bibfield  {journal} {\bibinfo  {journal} {PNAS}\ }\textbf {\bibinfo {volume} {111}},\ \bibinfo {pages} {6900} (\bibinfo {year} {2014})}\BibitemShut {NoStop}%
\bibitem [{\citenamefont {Leppert}\ \emph {et~al.}(2016)\citenamefont {Leppert}, \citenamefont {Reyes-Lillo},\ and\ \citenamefont {Neaton}}]{leppert2016electric}%
  \BibitemOpen
  \bibfield  {author} {\bibinfo {author} {\bibfnamefont {L.}~\bibnamefont {Leppert}}, \bibinfo {author} {\bibfnamefont {S.~E.}\ \bibnamefont {Reyes-Lillo}},\ and\ \bibinfo {author} {\bibfnamefont {J.~B.}\ \bibnamefont {Neaton}},\ }\bibfield  {title} {\bibinfo {title} {Electric field-and strain-induced rashba effect in hybrid halide perovskites},\ }\href@noop {} {\bibfield  {journal} {\bibinfo  {journal} {J. Phys. Chem. Lett.}\ }\textbf {\bibinfo {volume} {7}},\ \bibinfo {pages} {3683} (\bibinfo {year} {2016})}\BibitemShut {NoStop}%
\bibitem [{\citenamefont {Phuong}\ \emph {et~al.}(2016)\citenamefont {Phuong}, \citenamefont {Nakaike}, \citenamefont {Wakamiya},\ and\ \citenamefont {Kanemitsu}}]{phuong2016free}%
  \BibitemOpen
  \bibfield  {author} {\bibinfo {author} {\bibfnamefont {L.~Q.}\ \bibnamefont {Phuong}}, \bibinfo {author} {\bibfnamefont {Y.}~\bibnamefont {Nakaike}}, \bibinfo {author} {\bibfnamefont {A.}~\bibnamefont {Wakamiya}},\ and\ \bibinfo {author} {\bibfnamefont {Y.}~\bibnamefont {Kanemitsu}},\ }\bibfield  {title} {\bibinfo {title} {Free excitons and exciton--phonon coupling in ch3nh3pbi3 single crystals revealed by photocurrent and photoluminescence measurements at low temperatures},\ }\href@noop {} {\bibfield  {journal} {\bibinfo  {journal} {J. Phys. Chem. Lett.}\ }\textbf {\bibinfo {volume} {7}},\ \bibinfo {pages} {4905} (\bibinfo {year} {2016})}\BibitemShut {NoStop}%
\bibitem [{li2(2016)}]{li2016circular}%
  \BibitemOpen
  \bibfield  {title} {\bibinfo {title} {{Circular photogalvanic effect in organometal halide perovskite CH3NH3PbI3}, author={Li, Junwen and Haney, Paul M}},\ }\href@noop {} {\bibfield  {journal} {\bibinfo  {journal} {Appl. Phys. Lett.}\ }\textbf {\bibinfo {volume} {109}} (\bibinfo {year} {2016})}\BibitemShut {NoStop}%
\bibitem [{\citenamefont {Miyata}\ \emph {et~al.}(2015)\citenamefont {Miyata}, \citenamefont {Mitioglu}, \citenamefont {Plochocka}, \citenamefont {Portugall}, \citenamefont {Wang}, \citenamefont {Stranks}, \citenamefont {Snaith},\ and\ \citenamefont {Nicholas}}]{miyata2015direct}%
  \BibitemOpen
  \bibfield  {author} {\bibinfo {author} {\bibfnamefont {A.}~\bibnamefont {Miyata}}, \bibinfo {author} {\bibfnamefont {A.}~\bibnamefont {Mitioglu}}, \bibinfo {author} {\bibfnamefont {P.}~\bibnamefont {Plochocka}}, \bibinfo {author} {\bibfnamefont {O.}~\bibnamefont {Portugall}}, \bibinfo {author} {\bibfnamefont {J.~T.-W.}\ \bibnamefont {Wang}}, \bibinfo {author} {\bibfnamefont {S.~D.}\ \bibnamefont {Stranks}}, \bibinfo {author} {\bibfnamefont {H.~J.}\ \bibnamefont {Snaith}},\ and\ \bibinfo {author} {\bibfnamefont {R.~J.}\ \bibnamefont {Nicholas}},\ }\bibfield  {title} {\bibinfo {title} {Direct measurement of the exciton binding energy and effective masses for charge carriers in organic-inorganic tri-halide perovskites},\ }\href@noop {} {\bibfield  {journal} {\bibinfo  {journal} {Nat. Phys.}\ }\textbf {\bibinfo {volume} {11}},\ \bibinfo {pages} {582} (\bibinfo {year} {2015})}\BibitemShut {NoStop}%
\bibitem [{\citenamefont {Wang}\ \emph {et~al.}(2015)\citenamefont {Wang}, \citenamefont {Sumpter}, \citenamefont {Huang}, \citenamefont {Zhang}, \citenamefont {Liu}, \citenamefont {Yang},\ and\ \citenamefont {Zhao}}]{wang2015}%
  \BibitemOpen
  \bibfield  {author} {\bibinfo {author} {\bibfnamefont {Y.}~\bibnamefont {Wang}}, \bibinfo {author} {\bibfnamefont {B.~G.}\ \bibnamefont {Sumpter}}, \bibinfo {author} {\bibfnamefont {J.}~\bibnamefont {Huang}}, \bibinfo {author} {\bibfnamefont {H.}~\bibnamefont {Zhang}}, \bibinfo {author} {\bibfnamefont {P.}~\bibnamefont {Liu}}, \bibinfo {author} {\bibfnamefont {H.}~\bibnamefont {Yang}},\ and\ \bibinfo {author} {\bibfnamefont {H.}~\bibnamefont {Zhao}},\ }\bibfield  {title} {\bibinfo {title} {{Density Functional Studies of Stoichiometric Surfaces of Orthorhombic Hybrid Perovskite CH3NH3PbI3}},\ }\href {https://doi.org/10.1021/jp511123s} {\bibfield  {journal} {\bibinfo  {journal} {The Journal of Physical Chemistry C}\ }\textbf {\bibinfo {volume} {119}},\ \bibinfo {pages} {1136} (\bibinfo {year} {2015})}\BibitemShut {NoStop}%
\bibitem [{\citenamefont {Le}\ and\ \citenamefont {Sun}(2021)}]{Cle2021}%
  \BibitemOpen
  \bibfield  {author} {\bibinfo {author} {\bibfnamefont {C.}~\bibnamefont {Le}}\ and\ \bibinfo {author} {\bibfnamefont {Y.}~\bibnamefont {Sun}},\ }\bibfield  {title} {\bibinfo {title} {Topology and symmetry of circular photogalvanic effect in the chiral multifold semimetals: a review},\ }\href {https://doi.org/10.1088/1361-648X/ac2928} {\bibfield  {journal} {\bibinfo  {journal} {J. Phys. Condens. Matter}\ }\textbf {\bibinfo {volume} {33}},\ \bibinfo {pages} {3003} (\bibinfo {year} {2021})}\BibitemShut {NoStop}%
\bibitem [{\citenamefont {Leguy}\ \emph {et~al.}(2015)\citenamefont {Leguy}, \citenamefont {Frost}, \citenamefont {McMahon}, \citenamefont {Sakai}, \citenamefont {Kockelmann}, \citenamefont {Law}, \citenamefont {Li}, \citenamefont {Foglia}, \citenamefont {Walsh}, \citenamefont {O’regan} \emph {et~al.}}]{leguy2015dynamics}%
  \BibitemOpen
  \bibfield  {author} {\bibinfo {author} {\bibfnamefont {A.~M.}\ \bibnamefont {Leguy}}, \bibinfo {author} {\bibfnamefont {J.~M.}\ \bibnamefont {Frost}}, \bibinfo {author} {\bibfnamefont {A.~P.}\ \bibnamefont {McMahon}}, \bibinfo {author} {\bibfnamefont {V.~G.}\ \bibnamefont {Sakai}}, \bibinfo {author} {\bibfnamefont {W.}~\bibnamefont {Kockelmann}}, \bibinfo {author} {\bibfnamefont {C.}~\bibnamefont {Law}}, \bibinfo {author} {\bibfnamefont {X.}~\bibnamefont {Li}}, \bibinfo {author} {\bibfnamefont {F.}~\bibnamefont {Foglia}}, \bibinfo {author} {\bibfnamefont {A.}~\bibnamefont {Walsh}}, \bibinfo {author} {\bibfnamefont {B.~C.}\ \bibnamefont {O’regan}}, \emph {et~al.},\ }\bibfield  {title} {\bibinfo {title} {The dynamics of methylammonium ions in hybrid organic--inorganic perovskite solar cells},\ }\href@noop {} {\bibfield  {journal} {\bibinfo  {journal} {Nat. Commun.}\ }\textbf {\bibinfo {volume} {6}},\ \bibinfo {pages} {7124} (\bibinfo {year} {2015})}\BibitemShut {NoStop}%
\end{thebibliography}

\providecommand{\noopsort}[1]{}\providecommand{\singleletter}[1]{#1}%

\end{document}